\documentclass[graybox]{svmult}

\usepackage{natbib}
\usepackage{amssymb}
\usepackage{mathptmx}       
\usepackage{helvet}         
\usepackage{courier}        
\usepackage{type1cm}        
\usepackage{makeidx}         
\usepackage{graphicx}        
\usepackage{multicol}        
\usepackage[bottom]{footmisc}
\newcommand{\be}{\begin{equation}}
\newcommand{\ee}{\end{equation}}
\newcommand{\bea}{\begin{eqnarray}}
\newcommand{\eea}{\end{eqnarray}}

\makeindex      

\begin{document}

\title*{Magnetic Field Measurement with Ground State Alignment}
\author{Huirong Yan and A. Lazarian}
\institute{Huirong Yan \at KIAA, Peking University, Beijing 100871, China, \email{hryan@pku.edu.cn} 
\and A. Lazarian \at University of Wisconsin-Madison,
Astronomy Department, 475 N. Charter St., Madison, WI 53706, US, e-mail:
lazarian@astro.wisc.edu}
\maketitle

\abstract{Observational studies of magnetic fields are crucial. We introduce a process
"ground state alignment" as a new way to determine the  magnetic field
 direction in diffuse medium.  The alignment is due to anisotropic radiation impinging on the atom/ion. The consequence of the process is the polarization of spectral lines resulting from scattering and absorption from 
 aligned atomic/ionic species with fine or hyperfine structure. The magnetic field induces precession and realign the atom/ion and therefore the polarization of the emitted or absorbed radiation reflects the direction of the magnetic field. 
The atoms get aligned at their low levels and, as the life-time of the atoms/ions we deal with is long, the alignment induced by anisotropic radiation 
is susceptible to extremely weak magnetic fields ($1{\rm G}\gtrsim B\gtrsim 10^{-15}$G). In fact, the effects of atomic/ionic alignment were studied in the laboratory decades ago, mostly in relation to the maser research.  Recently, the atomic effect has been
already detected in observations from circumstellar medium and this is a harbinger of future extensive magnetic field studies. A unique feature of the atomic realignment is that they can reveal the 3D orientation of magnetic field. 
In this article, we shall review the basic physical processes involved in atomic realignment. We shall also discuss its applications to interplanetary, circumstellar and interstellar magnetic fields.  In addition, our research reveals that the polarization of the radiation arising from the transitions between fine and hyperfine states of the ground level can provide a unique diagnostics of magnetic fields in the Epoch of Reionization.}

\section{Introduction}

Astrophysical magnetic fields are ubiquitous and extremely important, especially in diffuse media, where their energy is comparable or exceed the energy of thermal gas. For instance, in diffuse interstellar medium (ISM), magnetic field pressure may exceed the thermal pressure by a factor of ten. In contrast, only a few techniques are available for the studies of magnetic field in diffuse medium and each of them has its own limitation. The Zeeman splitting can sample only relatively strong magnetic fields. in dense and cold clouds (see \cite{Crutcher:2010vn}). In most cases, only line of sight component of the field can be obtained. In some cases, the disentangling of the magnetic field and density fluctuations is nontrivial. For instance, the Faraday rotation is sensitive to the product of the electron density and the line-of-sight magnetic field (see \cite{Crutcher:2008ys}). Finally, all techniques have their area of applicability, e.g. polarization of the synchrotron emission traces the plane-of-sky magnetic fields of the galactic halo (see \cite{Beck:2011kx}). New promising statistical techniques can measure
the {\it average} direction of magnetic field using spectral lines fluctuations 
(\cite{Lazarian:2002uq}, \cite{Esquivel:2005qf}, \cite{Esquivel:2011bh}) or synchrotron intensity fluctuations \cite{Lazarian:2012fk}. 

The closest to the discussed technique of ground state alignment (henceforth GSA) are the techniques based on grain alignment and the Hanle effect.
It is well known  that the extinction and emission from aligned grains  reveal magnetic field direction perpendicular to the line of sight (see \cite{Hildebrand:2009zr} for a review).  In spite of the progress in understanding of grain alignment (see \cite{Lazarian07rev} for a review), the natural variations in grain shapes and compositions introduce uncertainties in the expected degree of polarization. In contrast, Hanle measurements were proposed for studies of circumstellar magnetic fields and require much higher magnetic fields \cite{:2004ff}. 

We should mention that all techniques suffer from the line-of-sight integration, which makes the tomography of magnetic fields difficult. As for the relative value, the most reliable is the Zeeman technique, but it is the technique that requires the strongest fields to study. In general, each technique is sensitive to magnetic fields in a particular environment  and the synergetic use of the technique is most advantageous. Obviously,  the addition of  a new techniques is a very unique and valuable development. 


Here we discuss a new promising technique to study magnetic fields in diffuse medium. As we discuss below, the physical foundations of these
technique can be traced back to the laboratory work on atomic alignment in the middle of the previous century (\cite{KASTLER-1950-234250}; \cite{refId0}; \cite{Hawkins:1953dz}; \cite{Hawkins:1955fv}; \cite{Cohen-Tannoudji:1969fk}; see \cite{YL_jqsrt} for details). Later
papers \cite{Varshalovich:1971mw} and \cite{Landolfi:1986lh} considered isolated individual cases of application of the aligned 
atoms mostly within toy models (see below for a brief review of the earlier development). Yan \& Lazarian (\citet{YLfine}, \cite{YLhyf}, \cite{YLHanle}, \cite{Yan:2009ys}) provided
detailed calculations of GSA for a number of atoms and through their study identified GSA as a very unique new technique applicable for studying
magnetic fields in a variety of environments, from circumstellar regions to the Early Universe. The emission and absorption lines ranging from radio to far UV were discussed. In particular, we identified new ways of study of magnetic fields using {\it absorption lines} (see \cite{YLfine}), radio lines arising from 
fine and hyperfine splitting \cite{YLHanle} and provided extensive calculations of expected polarization degree for a variety of ions and atoms most promising to trace magnetic fields in diffuse interstellar gas, protoplanetary nebula etc. 

The GSA technique as it stands now employs spectral-polarimetry and makes use of
the ability of atoms and ions to be aligned {\it in their ground state} by the external anisotropic radiation.
The aligned atoms interact with the astrophysical magnetic fields to get realigned. 

It is important to notice that the requirement for the
alignment in the ground state is the fine or hyperfine splitting of the ground state. The latter is true for
many species present in diffuse astrophysical environments.
Henceforth, we shall not
 distinguish atoms and ions and use word ``atoms'' dealing  with both species.
This technique can be used for 
interstellar\footnote{Here interstellar is understood in a 
general sense, which, for instance,
includes refection nebulae.}, and
 intergalactic studies as well as for studies of magnetic fields in
QSOs and other astrophysical objects.  

We would like to stress that the effect of ground-state atomic alignment is based on the well known physics. 
In fact, it has been known that atoms can be aligned through interactions with the anisotropic flux
of resonance emission (or optical pumping, see review \cite{Happer:1972ij} and references therein). Alignment is  understood here in terms of orientation
of the angular momentum vector
$\bf J$, if we use the language of classical mechanics. In quantum
terms  this means a difference in the population of sublevels corresponding to
projections of angular momentum to the quantization axis. There have been a lot of applications since the optical pumping was discovered by Kastler \cite{KASTLER-1950-234250}, ranging from atomic clocks, magnetometer, quantum optics and spin-polarized nuclei (see review by \citet{Budker:2007uq}; book by \citet{Cohen-Tannoudji:1969fk}). We will argue in
our review that similar fundamental changes  GSA can induce in terms of understanding magnetic fields in
diffuse media. 

In the review below we discuss three ways of using aligned atoms to trace magnetic field direction:
1) absorption lines
2) emission and fluorescent lines
3) emission and absorption lines related to transitions within splitting of the ground level

In addition, we shall discuss below how the information from the lines can be used to get the 3D structure of the magnetic fields, and, in particular cases, the intensity of magnetic field.

In terms of terminology, we will use "GSA" in the situations where magnetic field Larmor precession
is important and therefore the alignment reveals magnetic fields. Another possible term for the effect is
"atomic magnetic re-alignment", which stresses the nature of the effect that we discuss. However, whenever this
does not cause a confusion we prefer to use the term "GSA" in the analogy with the dust alignment
which is in most cases caused by radiation and reveals magnetic field due to dust Larmor precession in external magnetic fields 

In what follows, we describe the earlier work on atomic alignment in \S2, the basic idea of GSA in \S3. In \S4, we expatiate on absorption polarimetry, which is an exclusive tracer of GSA, we discuss polarimetry of both fine and hyperfine transitions, how to obtain from them 3D magnetic field, different regimes of pumping, circular polarization. Emission polarimetry in presented in \S5 and in \S6, we discuss another window of opportunity in IR and submillimetre based on the fine structure transitions within the aligned ground state. In \S7, the additional effect of GSA on abundance study is provided. In \S8, we put GSA in a context of broad view of radiative alignment processes in Astrophysics. In \S9, we focus on observational perspective and show a few synthetic observations with the input data on magnetic field from spacecraft measurements. Summary is provided in \S10.

\section{History on studies related to atomic alignment }

Unlike grain alignment, atomic alignment has been an established physical phenomena which has solid physical foundations and been studied and supported by numerous experiments (see review by \cite{Happer:1972ij}). The GSA was first proposed by Kastler \cite{KASTLER-1950-234250}, who received Nobel prize in 1966 for pointing out that absorption and scattering of resonant radiation, which is termed {\em optical pumping}, can induce imbalance on the ground state. Soon after that, the GSA was observed in experiments (\cite{refId0}; \cite{Hawkins:1953dz}). There have been a lot of applications since the optical pumping was proposed as a way of ordering the spin degrees of freedom, ranging from clocks, magnetometer, quantum optics and spin-polarized nuclei.

It is worth mentioning that atomic realignment in the presence of magnetic field was also
studied in laboratory in relation with early-day maser 
research (see \cite{Hawkins:1955fv}). Although our study in YL06 revealed that the mathematical treatment of the effect was not adequate in the original paper\footnote{Radiative pumping is much slower than magnetic mixing. Radiation was chosen as the quantization axis, nevertheless, which inevitably would lead to the nonzero coherence components. They were neglected in \cite{Hawkins:1955fv}, however. }  
, the importance of this pioneering study should not be underestimated.
The astrophysical application of the  GSA was first discussed in the interstellar medium context 
by \cite{Varshalovich:1968qc} for an atom with a hyperfine splitting. Varshalovich \cite{Varshalovich:1971mw} pointed out that  GSA 
can enable one to detect the direction of magnetic fields in the interstellar medium, and later in \cite{Varshalovich:1980fk} they proposed alignment of Sodium as a diagnostics of magnetic field in comet's head though the classical approach they used to describe the alignment in the presence of magnetic field is incorrect.  

Nearly 20 years after the work by Varshalovich, a case of emission of an idealized fine structure atom subject to a magnetic field and a beam of pumping radiation was conducted in \cite{Landolfi:1986lh}. However, in that case, a toy model
of a process, namely, an idealized 
two-level atom was considered. In addition, polarization of emission from this atom  was
discussed for a very restricted geometry of observations,
namely, the magnetic field is along the line of sight and both of these directions are perpendicular to the
 beam of incident light. 
This made it rather difficult to use this study as a tool for practical mapping of magnetic fields in various astrophysical environments.

The  GSA we deal with in this review should not be confused with the Hanle effect that solar researcher have extensively studied. While both effects are based on similar atomic physics and therefore share some of the quantum electrodynamic 
machinery for their calculations, the domain of applicability of the effects is very different. In particular, 
Hanle effect is depolarization and rotation of the polarization vector of the resonance scattered lines in the presence of a magnetic field, which happens when the magnetic splitting becomes comparable to the decay rate of the excited state of an atom. The research into emission line polarimetry resulted in important
change of the views on solar chromosphere (see \cite{Landi-DeglInnocenti:1983mi}, \cite{Landi-DeglInnocenti:1984kl}, \cite{Landi-DeglInnocenti:1998pi}, \cite{Stenflo:1997cq}, \cite{Trujillo-Bueno:1997fc}, \cite{Trujillo-Bueno:2002rq}). However, these studies correspond to a setting different from the one we consider in the case of  GSA. The latter is the weak field regime,  for which the Hanle effect is negligible. As we mentioned
earlier, in the  GSA regime the atoms/ions at ground level are repopulated due to magnetic precession. While Hanle
effect is prominent in the Solar case, it gets too weak for the environments of interstellar media, circumstellar regions and plasmas
of Early Universe. These are the areas where the  GSA effect is expected to be very important. 

The realignment happens if during the lifetime of an atomic state more than one Larmor precession happens. The
time scale of atomic precession scales as $0.011(5\mu G/B)$ s. As the life time
of the ground state is long typically (determined by absorption rate, see table 1), even extremely weak magnetic fields can be detected this way. On the
contrary, the typical application of the Hanle effect includes excited states with typical life-times of $A^{-1}\gtrsim 10^7$. 
Therefore, unlike GSA Hanle effect is used for studies of relatively strong magnetic fields, e.g.
magnetic fields of the stars (see \cite{Nordsieck:2005uq}). 

Full calculations of alignment of atom and ions in their ground or metastable state in the presence of magnetic field were done in Yan \& Lazarian (\cite{YLfine}, \cite{YLhyf}, \cite{YLHanle}) considers polarization of absorbed light arising from aligned atoms with fine structure, \cite{YLhyf} extends the treatment to emission and atoms with hyperfine, as well as, fine and hyperfine structure. \cite{YLHanle} addresses the issues of radio emission arising from the transitions between the sublevels of the ground state and extended the discussions to the domain of both stronger and weaker magnetic field when the Hanle and ground level Hanle effects are present.  

The polarization arising from  GSA is on its way of becoming an accepted tool for interstellar and circumstellar
studies. We see some advances in this direction. For instance,  polarization of absorption lines arising from
 GSA was predicted in YL06 and was detected for the polarization of H$\alpha$ absorption in \cite{Kuhn:2007vn} though they neglected the important realignment effect of magnetic field in their analysis. Focused on atomic fluorescence, Nordsieck \cite{Nordsieck:2008kx} discussed observational perspective using pilot spectroscopic observation of NGC as an example. We are sure that more detection of the predicted polarization will follow soon. Therefore we believe that the time is ripe to
discuss the status of the  GSA in the review in order to attract more attention of both observers and theorists to this promising effect.

\section{Basic physics of GSA}
\begin{figure*}[!t]
\includegraphics[%
  width=0.35\textwidth,
  height=0.18\textheight]{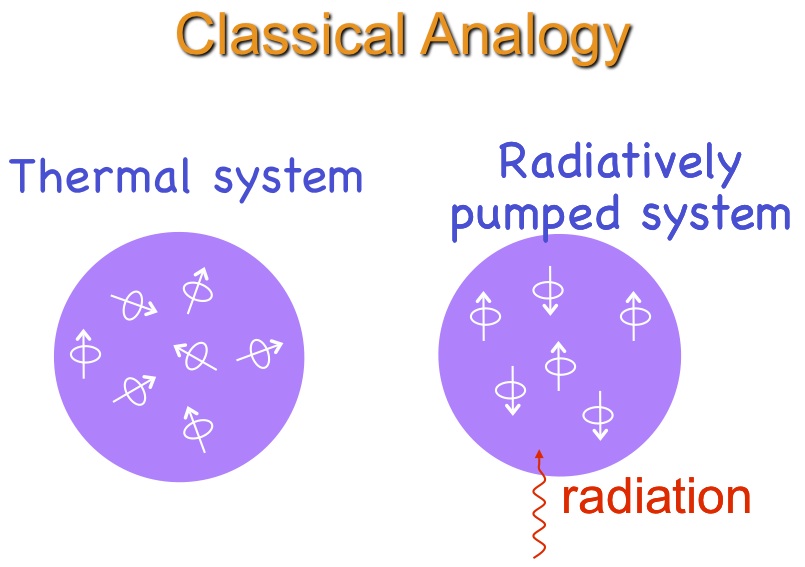}
 \includegraphics[%
  width=0.35\textwidth,
  height=0.2\textheight]{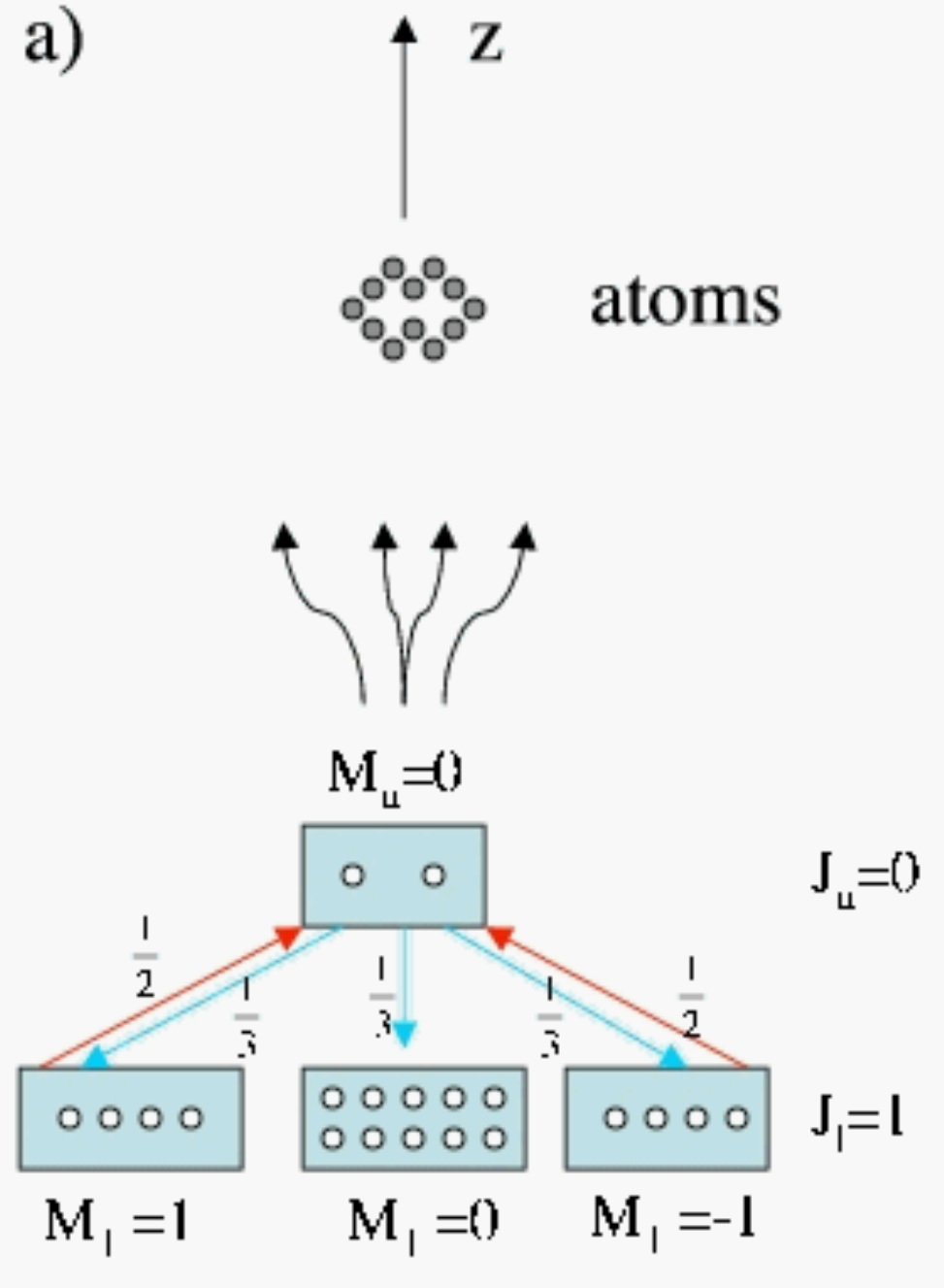}
  \includegraphics[%
  width=0.25\textwidth,
  height=0.2\textheight]{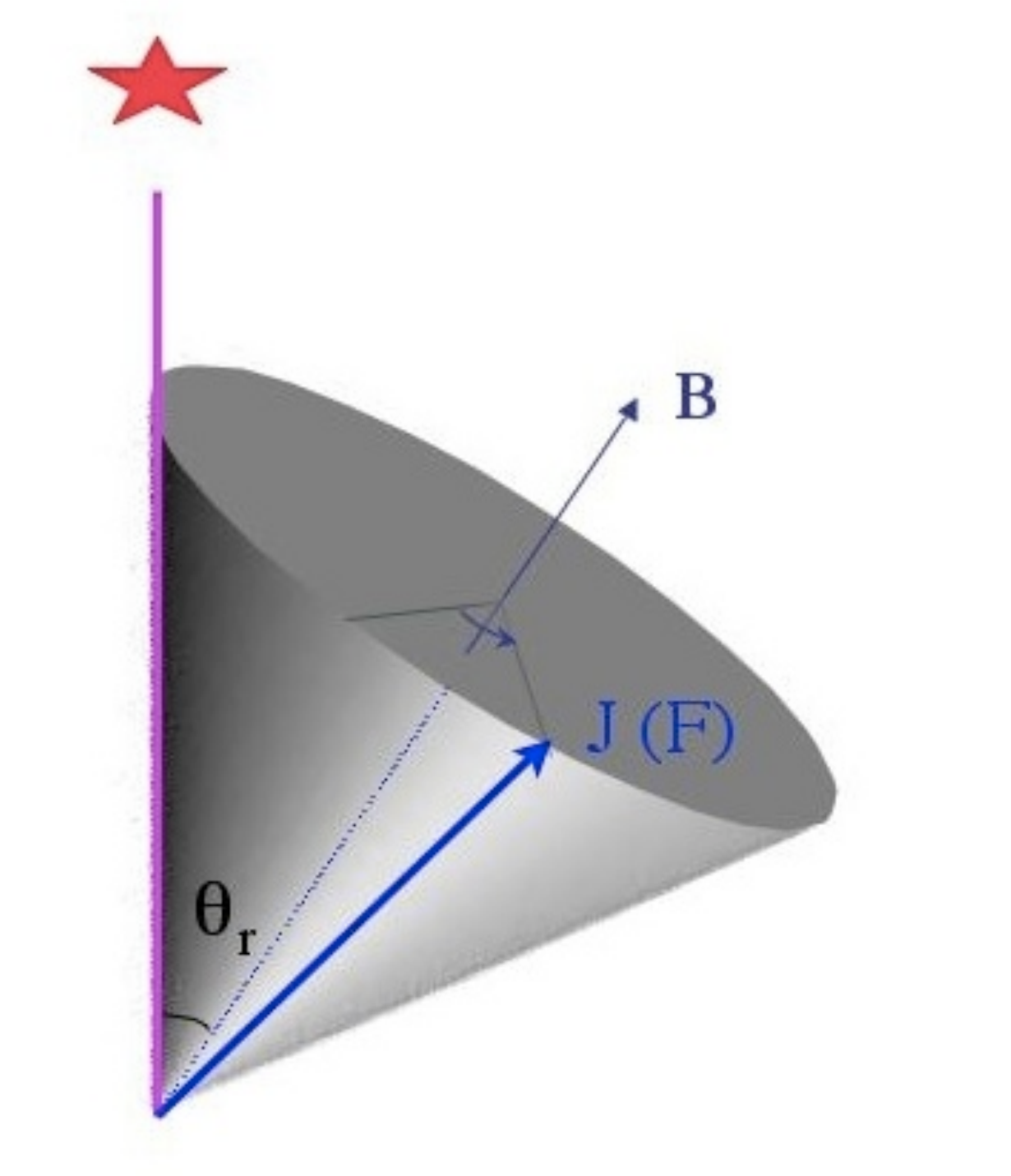}
\caption{{\em Upper left}: A carton illustrate classical analogy of the GSA induced by optical pumping; {\em Lower left}: A toy model to illustrate how atoms are aligned by anisotropic light.
Atoms accumulate in the ground sublevel $M=0$ as radiation removes atoms from the ground states $M=1$ and $M=-1$; {\em Upper right}: Typical astrophysical environment where the ground-state atomic alignment can happen. A pumping source deposits angular momentum to atoms in the direction of radiation and causes differential occupations on their ground states. {\em Lower right}: In a magnetized medium where the Larmor precession rate $\nu_L$ is larger than the photon arrival rate $\tau_R^{-1}$, however, atoms are realigned with respect to magnetic field. Atomic alignment is then determined by $\theta_r$, the angle between the magnetic field and the pumping source. The polarization of scattered line also depends on the direction of line of sight, $\theta$ and $\theta_0$. (From \cite{YLHanle})}
\label{nzplane}
\end{figure*}

The basic idea of the GSA is very
simple. The alignment is caused by
the anisotropic deposition of angular momentum from photons of {\it unpolarized} radiation. In typical
 astrophysical situations the radiation
flux is anisotropic\footnote{Modern theory of dust alignment, which is a very powerful way to study 
magnetic fields (see \cite{Lazarian07rev} and ref. therein) is also appealing to anisotropic radiation as the
cause of alignment.} (see Fig.\ref{nzplane}{\it right}). As the photon
spin is along the direction of its propagation, we expect that atoms
scattering the radiation from a light beam can aligned. Such an alignment happens in terms of 
the projections of angular momentum to the direction of the incoming light. To
study weak magnetic fields, one should use atoms that can be aligned in the ground state. For such atoms to be aligned, their ground state should have the non-zero angular momentum. Therefore fine (or hyperfine) structure is {\it necessary} for the alignment that we describe in the review.

Let us discuss a toy model that provides an intuitive  insight into  the 
physics of GSA.
Consider an atom with its ground state corresponding to the total angular momentum
$I=1$ and the upper state corresponding to the angular momentum $I=0$ \cite{Varshalovich:1971mw}
If the projection of the angular momentum to the direction of the
incident resonance photon beam is $M$, the upper state $M$ can
have values $-1$, $0$, and $1$, while for the upper state M=0 (see Fig.\ref{nzplane}{\it left}). 
The unpolarized
beam contains an equal number of left and right circularly polarized
photons whose projections on the beam direction are 1 and -1. Thus
absorption of the photons will induce transitions from the $M=-1$ and
$M=1$ sublevels. However, the decay from the upper state populates all
the three sublevels on ground state. As the result the atoms accumulate in the $M=0$ ground
sublevel from which no excitations are possible. Accordingly, the optical
properties of the media (e.g. absorption) would change.

The above toy model can also exemplify the
 role of collisions and magnetic field. Without collisions one may expect 
that all atoms
reside eventually at the sublevel of $M=0$. Collisions, however, redistribute
atoms to different sublevels. Nevertheless, as the randomization of the ground
state requires spin flips, it is less efficient than one might naively
imagine \cite{Hawkins:1955fv}. For instance, experimental study in \cite{Kastler:1957ys} suggests that
more than 10 collisions with electrons. are necessary to
destroy the aligned state of sodium. The reduced sensitivity of aligned
atoms to disorienting collisions makes the effect important for various
astrophysical environments.

Owing to the precession, the atoms with different projections of angular momentum will be mixed up. 
Magnetic mixing happens if the angular
momentum precession rate  is higher than the rate of the
excitation from the ground state, which is true for many astrophysical conditions, e.g., interplanetary medium, ISM, intergalactic medium, etc. 
As a result, angular momentum is redistributed among the atoms, and the alignment is altered according to the angle between the magnetic field and radiation field $\theta_r$ (see Fig.\ref{nzplane}{\em right}). This is the {\em classical} picture. 

In {\em quantum} picture, if magnetic precession is dominant, then the natural quantization axis will be the magnetic field, which in general is different from the symmetry axis of the radiation. The radiative pumping is to be seen coming from different directions according to the angle between the magnetic field and radiation field $\theta_r$, which results in different alignment.

The classical theory can give a qualitative interpretation which shall be utilized in this paper to provide an intuitive picture. Particularly for emission lines, both atoms and the radiation have to be described by the density matrices in order to obtain quantitative results. This is because there is coherence among different magnetic sublevels on the upper state\footnote{In quantum physics, quantum coherence means that subatomic particles are able to cooperate. These subatomic waves or particles not only know about each other, but are also highly interlinked by bands of shared electromagnetic fields so that they can communicate with each other.}.

Our simple considerations above indicate that, in order to be aligned, first, atoms should have enough 
degrees of freedom: namely, the quantum angular momentum number must be 
$\ge 1$. Second, the incident flux must be anisotropic. 
Moreover, the collisional rate should not be too high. While the latter
requires special laboratory conditions, it is applicable to many astrophysical environments such as the outer
layers of stellar atmospheres, the interplanetary,
interstellar, and intergalactic medium. 

\subsection{Timescales and different regimes of pumping}

In terms of practical magnetic field studies, the variety of available species is important in many aspects. One of them is a possibility of
getting additional information about environments. Let us illustrate this by considering
the various rates (see Table \ref{difftime}) involved. Those 
are 1) the rate of the Larmor precession, $\nu_L$, 2) the rate of the optical pumping, $\tau_R^{-1}$, 3) the rate of collisional randomization, $\tau_c^{-1}$,
4) the rate of the transition within ground state, $\tau^{-1}_T$.  
In many cases $\nu_L>\tau_R^{-1}>\tau_c^{-1}, \tau_T^{-1}$. 
Other relations are possible, however. If $\tau_T^{-1}>\tau_R^{-1}$, the transitions within the sublevels of ground state need to be taken into account and relative distribution among them will be modified (see YL06,c). Since emission is spherically symmetric, the angular momentum in the atomic system is preserved and thus alignment persists in this case. In the case $\nu_L<\tau_R^{-1}$, the magnetic field does not affect the atomic occupations and atoms are aligned with respect to the direction of radiation. From the expressions in Table~\ref{difftime}, we see, for instance, that magnetic field can realign CII only at a distance $r\gtrsim7.7$Au from an O star if the magnetic field strength $\sim 5\mu$G.

If the Larmor precession rate $\nu_L$ is comparable to any of the other rates,
the atomic line polarization becomes sensitive to the strength of the magnetic field. In these situations, it is possible to get information about the {\it magnitude} of magnetic
field. 
\begin{table*}[!t]
\begin{tabular}{clcc}
\hline
$\nu_L$(s$^{-1}$)&Larmor precession frequency& $\frac{eB}{m_ec}$&$88(B/5\mu$ G) \\
$\tau_R^{-1}$(s$^{-1}$)&radiative pumping rate&$B_{J_lJ_u}I$& $7.4\times 10^{5}\left(\frac{R_*}{r}\right)^2$\\
$\tau_T^{-1}$(s$^{-1}$)&emission rate within ground state &$A_m$&2.3$\times 10^{-6}$\\
$\tau_c^{-1}$(s$^{-1}$)&collisional transition rate&max($f_{kj},f_{sf}$) &$6.4\left(\frac{n_e}{0.1{\rm cm}^{-3}}\sqrt{\frac{8000{\rm K}}{T}}\right)\times 10^{-9}$\\
\hline
\end{tabular}
\caption{{\small Relevant rates for GSA.   $A_m$ is the magnetic dipole emission rate for transitions among J levels of the ground state of an atom. $f_{kj}$ is the inelastic collisional transition rates within ground state due to collisions with electrons or hydrogens, and $f_{sp}$ is the spin flip rate due to Van der Waals collisions. In the last row, example values for C II are given. $\tau_R^{-1}$ is calculated for an O type star, where $R_*$ is the radius of the star radius and r is the distance to the star. (From \cite{YLfine})}}
\label{difftime}
\end{table*}

\begin{figure}[!t]
  \includegraphics[width=0.8\columnwidth]{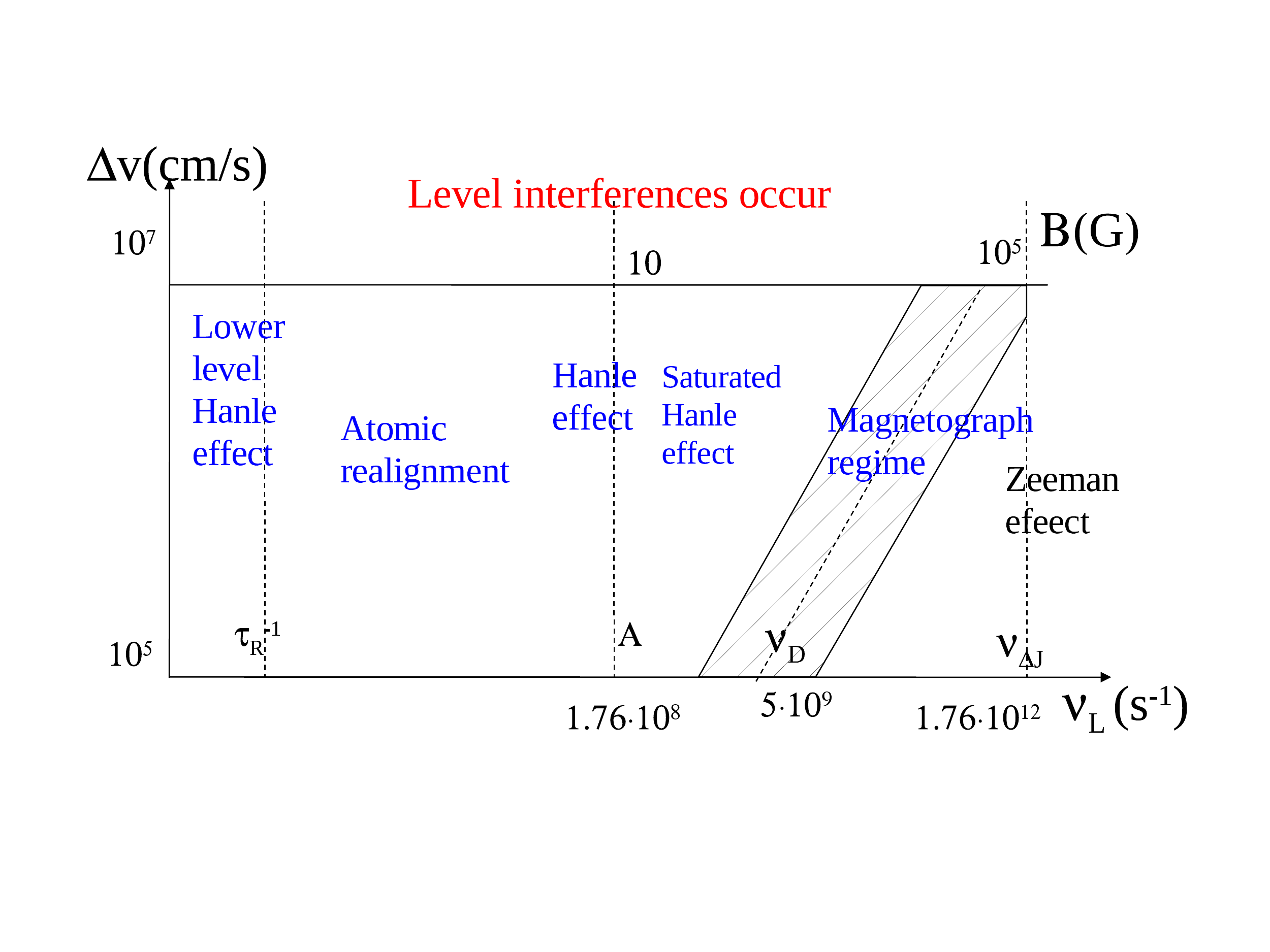}
  \caption{ Different regimes divided according to the strength of magnetic field and the Doppler line width. Atomic realignment is applicable to weak field ($<1G$) in diffuse medium. Level interferences are negligible unless the medium is substantially turbulent ($\delta v\gtrsim$ 100km/s) and the corresponding Doppler line width becomes comparable to the fine level splitting $\nu_{\Delta J}$. For strong magnetic field, Zeeman effect dominates. When magnetic splitting becomes comparable to the Doppler width, $\sigma$ and $\pi$ components (note: we remind the reader that $\sigma$ is the circular polarization and $\pi$ represents the linear polarization.) can still distinguish themselves through polarization, this is the magnetograph regime; Hanle effect is dominant if Larmor period is comparable to the lifetime of excited level $\nu_L^{-1}\sim A^{-1}$; similarly, for ground Hanle effect, it requires Larmor splitting to be of the order of photon pumping rate; for weak magnetic field ($<1G$) in diffuse medium, however, GSA is the main effect provided that $\nu_L=17.6(B/\mu G){\rm s}^{-1}>\tau_R^{-1}$. (From \cite{YLfine})}
  \label{regimes}
\end{figure}

Fig.\ref{regimes} illustrates the regime of magnetic field strength where atomic realignment applies. Atoms are aligned by the anisotropic radiation at a rate of $\tau_R^{-1}$. Magnetic precession will realign the atoms in their ground state if the Larmor precession rate $\nu_L>\tau_R^{-1}$. In contrast, if the magnetic field gets stronger so that Larmor frequency becomes comparable to the line-width of the upper level, the upper level occupation, especially coherence is modified directly by magnetic field, this is the domain of Hanle effect, which has been extensively discussed for studies of solar magnetic field (see \cite{:2004ff} and references therein). When the magnetic splitting becomes comparable to the Doppler line width $\nu_D$, polarization appears, this is the ``magnetograph regime'' \cite{Landi-DeglInnocenti:1983mi}. For magnetic splitting $\nu_L\gg \nu_D$, the energy separation is enough to be resolved, and the magnetic field can be deduced directly from line splitting in this case. If the medium is strongly turbulent with $\delta v\sim 100$km/s (so that the Doppler line width is comparable to the level separations $\nu_D\sim \nu_{\Delta J}$), interferences occur among these levels and should be taken into account.

Long-lived alignable metastable states that are present for
some atomic species between upper and lower states may act as
proxies of ground states.  Absorptions from these metastable levels
 can also be used as diagnostics for magnetic field therefore.

The variety of species that are subject to the ground state or metastable state alignment render the GSA technique with really unique capabilities. Different species are expected to be aligned when the
conditions for their existence and their alignment are satisfied. This allows to study the 3D distribution of magnetic
fields, rather than line-average magnetic fields, which are available through most of the alternative techniques. 

\begin{figure}
\includegraphics[%
  width=0.38\textwidth,
  height=0.25\textheight]{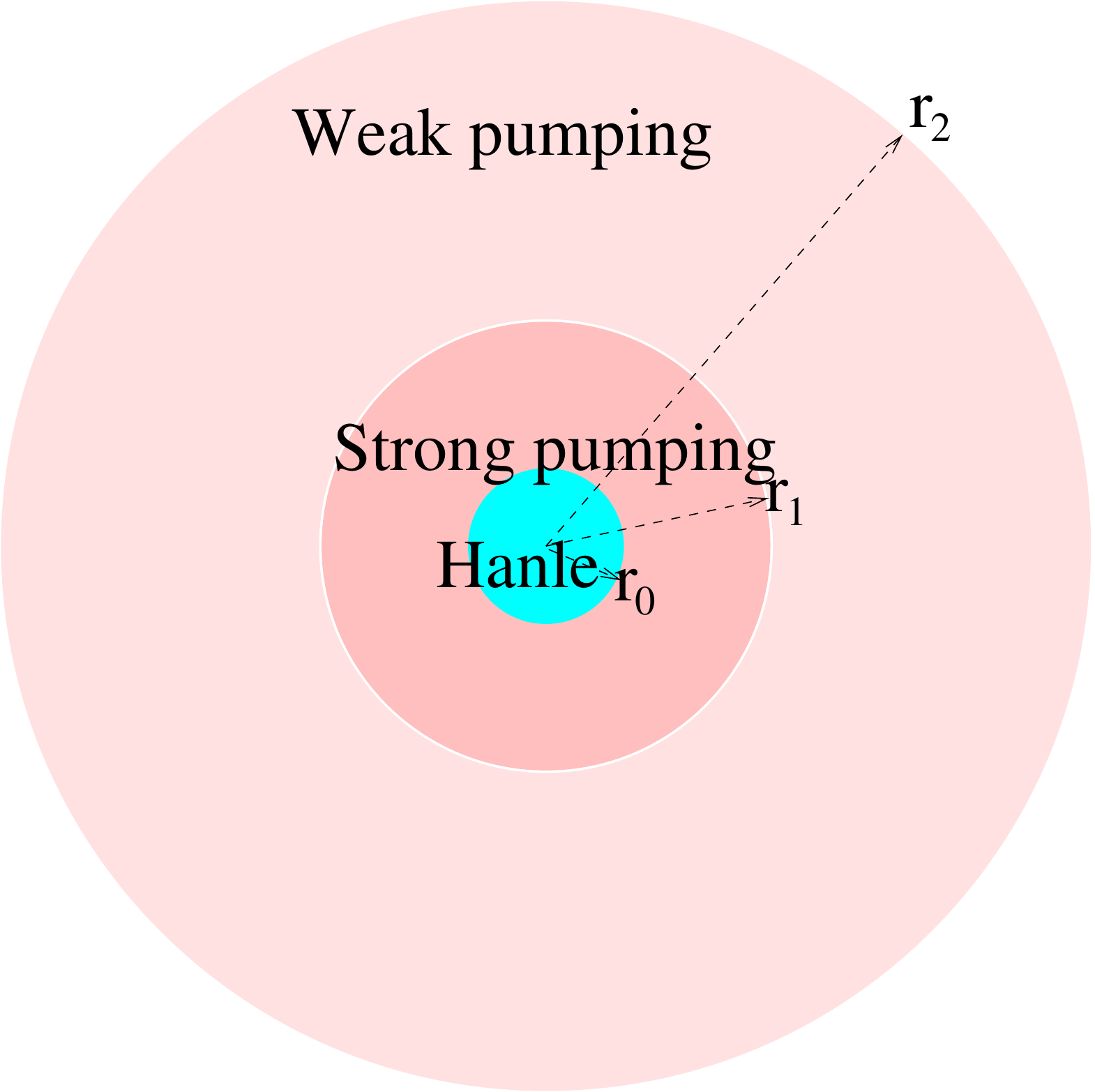}
\includegraphics[%
  width=0.65\textwidth,
  height=0.35\textheight]{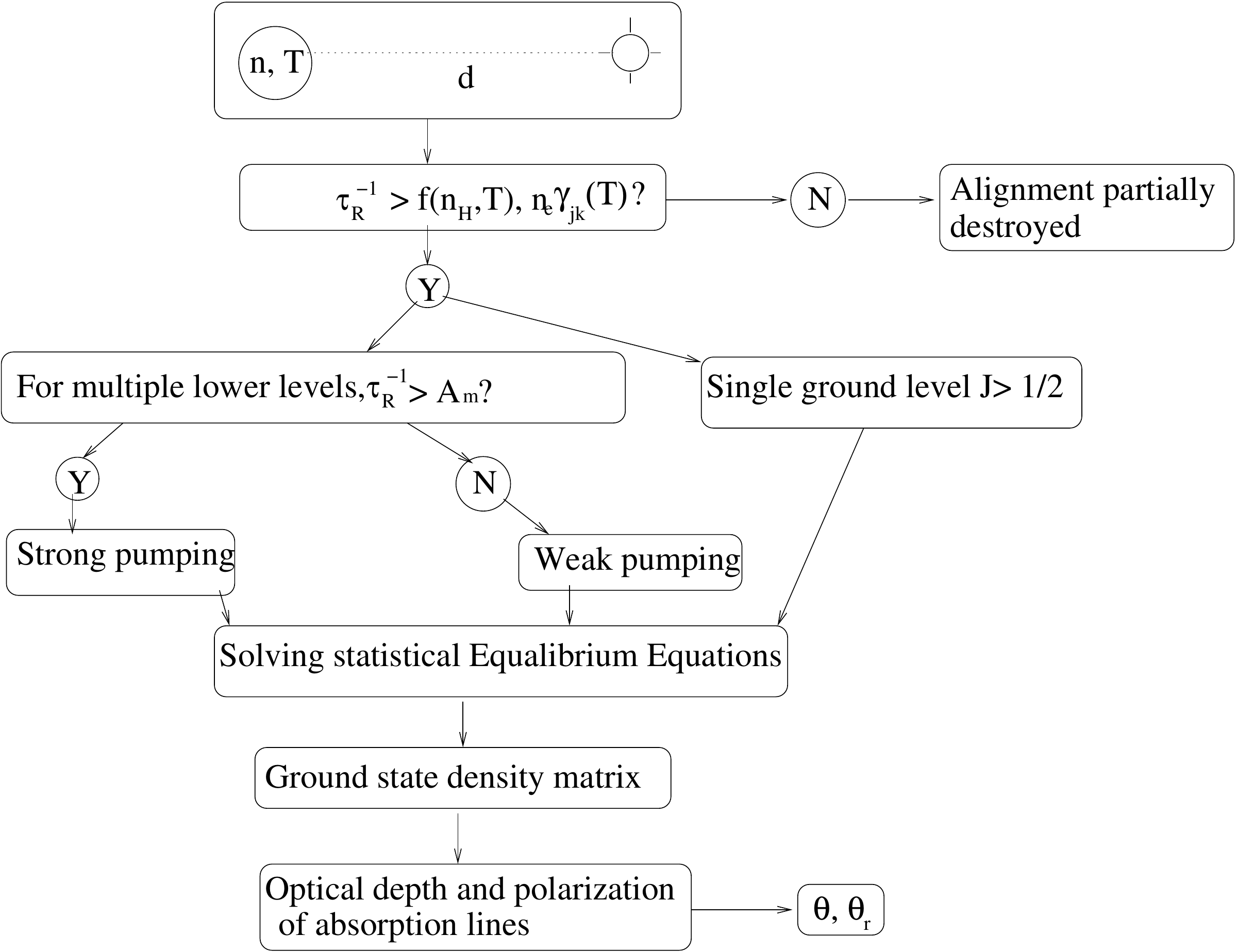}
\caption{{\it left}: a carton illustrating how the atomic pumping changes with distance around a radiation source. For circumsteller region, magnetic field is strong, such that the Hanle effect, which requires $\nu_L\sim A$, dominates. Atomic alignment applies to the much more distant interstellar medium, within $r_2$, which is defined as the radius where the optical pumping rate $\tau_R^{-1}$ is higher than collisional rate. Inside $r_2$, it can be further divided into two regimes: strong pumping and weak pumping, demarcated at $r_1$ (see Eq.\ref{demarcator}); {\it Right}: whether and how atoms are aligned depends on their intrinsic properties (transitional probabilities and structures) and the physical conditions: density n, temperature T and the averaged radiation intensity from the source $I_*$. If the pumping rate $\tau_R^{-1}$ is less than collisional rates, alignment is partially destroyed. Then for atoms with multiple lower levels, depending on the comparison between the pumping rate and the magnetic dipole radiation rate among the lower levels, the atoms are aligned differently. In the strong pumping case, all the alignable lower levels are aligned; on the contrary, only the ground level can be aligned in the case of weak pumping. From \cite{YLfine}.}
\label{diagram}
\end{figure}

Most atoms have sublevels on the ground state, among which there are magnetic dipole transitions. Although its transition probability is very low, it can be comparable to the optical pumping rate in regions far from any radiation source. Depending on how far away the radiation source is, there can be two regimes divided by the boundary where the magnetic  dipole radiation rate $A_m$ is equal to the pumping rate $\tau_R^{-1}$. Inside the boundary, the optical pumping rate is much larger than the M1 transition rate $A_m$ so that we can ignore the magnetic dipole radiation as a first order approximation. Further out, the magnetic dipole emission is faster than optical pumping so that it can be assumed that most atoms reside in the lowest energy level of the ground state and alignment can only be accommodated on this level. 
The two regimes are demarcated at $r_1$ (see Fig.\ref{diagram} {\it left}), where $r_1$ is defined by 
\be 
\tau_R^{-1}=B_{J_lJ_u}I_{BB}(R_*/r_1)^2=A_m,
\label{demarcator}
\ee 
and $I_{BB}, R_*$ are the intensity and radius of the pumping source. For different radiation sources, the distance to the boundary differs. For an O type star, the distance would be $\sim$0.1pc for species like C II, Si II, while for a shell star ($T_{eff}=15,000$K), it is as close as to $\sim 0.003-0.01$ pc. For the species like, C II, Si II  the lowest level is not alignable with $J_l=1/2$, and thus the alignment is absent outside the radius $r_1$, which ensures that observations constrain the magnetic field topology within this radius.

\section{Polarization of absorption lines}

The use of absorption lines to study magnetic fields with aligned atoms
was first suggested by Yan \& Lazarian (2006) \cite{YLfine}. Below we briefly outline the main ideas of the
proposed techniques.

When atomic species are aligned on their ground state, the corresponding absorption from the state will be polarized as a result of the differential absorption parallel and perpendicular to the direction of alignment.  The general expression for finite optical depth would be  
\bea
I&=&(I_0+Q_0)e^{-\tau(1+\eta_1/\eta_0)}+(I_0-Q_0)e^{-\tau(1-\eta_1/\eta_0)},\nonumber\\
Q&=&(I_0+Q_0)e^{-\tau(1+\eta_1/\eta_0)}-(I_0-Q_0)e^{-\tau(1-\eta_1/\eta_0)},\nonumber\\
U&=&U_0e^{-\tau}, V=V_0e^{-\tau},\eea
where $I_0, Q_0, U_0$ are the Stokes parameters of the background radiation, which can be from a weak background source or the pumping source itself. $d$ refers to the thickness of medium. $\eta_0, \eta_1,\eta_2$ are the corresponding absorption coefficients. In the case of unpolarized background radiation and thin optical depth, the degree of linear polarization is given by

\be 
P=\frac{Q}{I}=\frac{e^{-(\eta_0+\eta_1)d}-e^{-(\eta_0-\eta_1)d}}{e^{-(\eta_0+\eta_1)d}+e^{-(\eta_0-\eta_1)d}} \approx -\tau \frac{\eta_1}{\eta_0}
\ee

The polarization in this case has a simple relation given by
\be
\frac{P}{\tau}=\frac{Q}{I\eta_0d}\simeq\frac{-\eta_1}{\eta_0}=\frac{1.5\sigma^2_0(J_l, \theta_r)\sin^2\theta w^2_{J_lJ_u}}{\sqrt{2}+\sigma^2_0(J_l, \theta_r)(1-1.5\sin^2\theta)w^2_{J_lJ_u}}.
\label{absorb}
\ee
with the alignment parameter $\sigma_0^2\equiv \frac{\rho^2_0}{\rho^0_0}$, the normalized dipole component of density matrix of ground state, which quantifies the degree of alignment. For instance, for a state of angular momentum 1, the definition of $\rho^2_0=[\rho(1,1)-2\rho(1,0)+\rho(1,-1)]$. The generic definition can be found in \cite{Fano:1957, DYakonov:1965}.
The sign of it gives the direction of alignment. Since magnetic field is the quantization axis, a positive alignment parameter means that the alignment is parallel to the magnetic field and a negative sign means the alignment is perpendicular to the magnetic field. $\theta$ is the angle between the line of sight and magnetic field. $w^2_{J_lJ_u}$ defined below, is a parameter determined by the atomic structure
\be
w^K_{J_lJ_u}\equiv\left\{\begin{array}{ccc}1 & 1 & K\\J_l&J_l& J_u\end{array}\right\}/\left\{\begin{array}{ccc}1 & 1 & 0\\J_l&J_l& J_u\end{array}\right\}.
\label{w2}
\ee
The values of $w^2_{JJ'}$ for different pairs of $J,J'$ are listed in Table~\ref{ch3t3}. We see that it totally depends on the  sign of  $w^2_{JJ'}$, whether the alignment and polarization are either parallel or orthogonal. Once we detect the direction of polarization of some absorption line, we immediately know the direction of alignment.
{
\begin{table}
\begin{tabular}{||c|ccc|ccc|ccc||}
\hline
\hline
 $J$&
 \multicolumn{3}{c|}{1}&
 \multicolumn{3}{c|}{3/2}&
 \multicolumn{3}{c||}{2}
  \tabularnewline
\hline 
 $J'$&
 0&1&2& 1/2&3/2&5/2 &1&2&3\\
\hline
 $w^2_{JJ'}$&
 1&-0.5&0.1& 0.7071&-0.5657&0.1414 &0.5916&-0.5916&0.1690\\
 \hline
 \hline 
 $J$&
 \multicolumn{3}{c|}{5/2}&
 \multicolumn{3}{c|}{3}&
 \multicolumn{3}{c||}{7/2}
  \tabularnewline
\hline 
 $J'$&
 3/2&5/2&7/2 &2&3&4& 5/2&7/2&9/2\\
\hline
 $w^2_{JJ'}$&
 0.5292&-0.6047&0.1890& 0.4899&-0.6124&0.2041& 0.4629&-0.6172&0.2160\\
 \hline
 \hline
 $J$&
 \multicolumn{3}{c|}{4}&
 \multicolumn{3}{c|}{9/2}&
 \multicolumn{3}{c||}{5}
  \tabularnewline
\hline 
 $J'$&
 3&4&5& 7/2&9/2&11/2 &4&5&6\\
\hline
 $w^2_{JJ'}$&
 0.4432&-0.6205&0.2256& 0.4282&-0.6228&0.2335& 0.4163&-0.6245&0.2402\\
 \hline
\hline
\end{tabular}
\caption{Numerical values of $w^2_{JJ'}$. $J$ is the J value of the initial level and $J'$ is that of the final level.}
\label{ch3t3}
\end{table}
}
The alignment is either parallel or perpendicular to the direction of symmetry axis of pumping radiation in the absence of magnetic field. Real astrophysical fluid though is magnetized, and the Larmor precession period is usually larger than the radiative pumping rate unless it is very close to the radiation source as we pointed out earlier. In this case, realignment happens and atomic species can be aligned either parallel or perpendicular to the local magnetic field. The switch between the two cases is always at the Van Vleck angle $\theta_r = 54.7^\circ,\, 180^\circ-54.7^\circ$, where $\theta_r$ is the angle between the magnetic field and radiation. As the result the polarization of the absorption line also changes according to Eq.(\ref{absorb}). {\it This turnoff at the Van Vleck angle is a generic feature regardless of the specific atomic species as long as the background source is unpolarized and it is in the atomic realignment regime.} In practice, this means that once we detect any polarization in absorption line, we get immediate information of the direction of magnetic field in the plane of sky within 90 degree degeneracy. If we have two measurable,  then according to eq.(\ref{absorb}) both $\theta_r, \theta$ can be determined. With $\theta_r$ known, the 90 degree degeneracy in the pictorial plane can be removed and we can get 3D information of magnetic field. 

\begin{table*}
{\footnotesize \begin{tabular}{ccccc}
\hline\hline
Species&Ground state&excited state&Wavelength(\AA)&$P_{max}$\\[0.5ex]
\hline
S II&$4S^o_{3/2}$&
$4P_{1/2,3/2,5/2}$& 1250-1260&12\%($3/2\rightarrow 1/2$)\\
\hline
Cr I&
$a7S_3$&
$7P^o_{2,3,4}$&
3580-3606&5\%($3\rightarrow 2$)\\
\hline
 C II&
$2P^o_{1/2,3/2}$&
$2S_{1/2}$,$2P_{1/2,3/2}$,$2D_{3/2,5/2}$&
1036.3-1335.7&15\%($3/2\rightarrow 1/2$)\\
Si II&&&989.9-1533.4&7\%($3/2\rightarrow 1/2$)\\
\hline
 O I&$3P_{2,1,0}$
&$3S_1$, $3D_{1,2,3}$&
911-1302.2&29\%($2\rightarrow 2$)\\
S I&&$3S_1$,$3P^o_{0,1,2}$, $3D_{1,2,3}$&1205-1826&22\%($1\rightarrow 0$)\\
\hline
 C I&&&
 1115-1657&18\%($1\rightarrow 0$)\\
Si I&$3P_{0,1,2}$&
$3P^o_{0,1,2}$,$3D^o_{1,2,3}$&
1695-2529&20\%($2\rightarrow 1$)\\
S III &&&$$1012-1202&24.5\%($2\rightarrow 1$)\\
\hline
&&$z4G^o_{5/2}$&3384.74&-0.7\%\\
{TiII}&$a4F_{3/2}$&$z4F^o_{5/2}$&3230.13&-0.7\%\\
&&$z4F^o_{3/2}$&3242.93&2.9\%\\
&&$z4D^o_{3/2}$&3067.25&2.9\%\\
&&$z4D^o_{1/2}$&3073.88&7.3\%\\
\hline
\hline
\end{tabular}}
\caption{Absorption lines of selected alignable atomic species and corresponding transitions. Note only lines above $912 \AA$ are listed. Data are taken from the Atomic Line List http://www.pa.uky.edu/$\sim$peter/atomic/ and the NIST Atomic Spectra Database. The last column gives the maximum polarizations and its corresponding transitions. For those species with multiple lower levels, the polarizations are calculated for shell star ($T_{eff}=15,000$K) in the strong pumping regime; in the weak pumping regime, the maximum polarizations are 19\% for OI transition ($2\rightarrow 2$), and 9\% for SI transition ($2\rightarrow 2$).}
\label{species}
\end{table*}

\begin{figure}
\includegraphics[%
  width=0.4\textwidth,
  height=0.25\textheight]{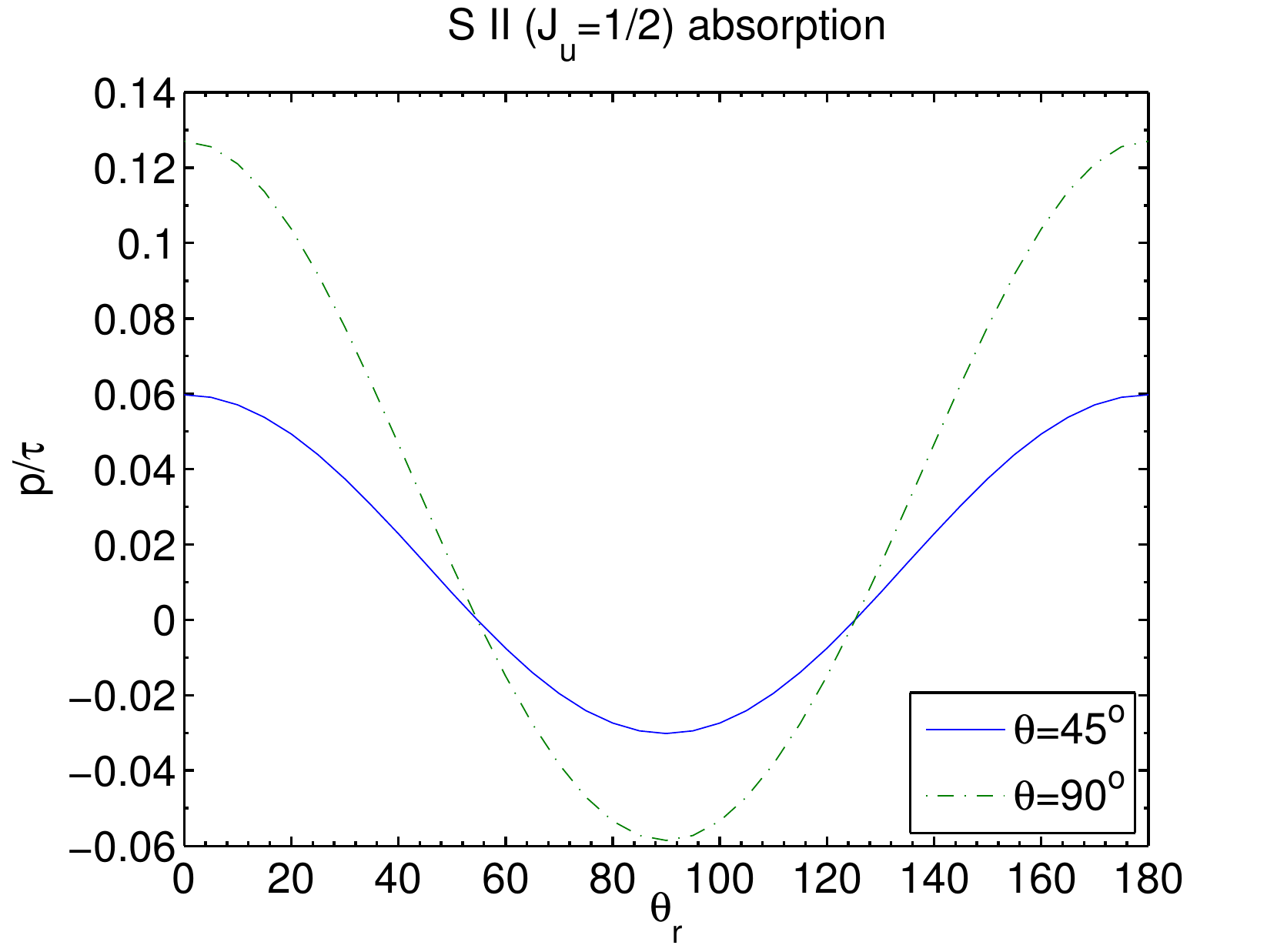}
\includegraphics[%
  width=0.4\textwidth,
  height=0.25\textheight]{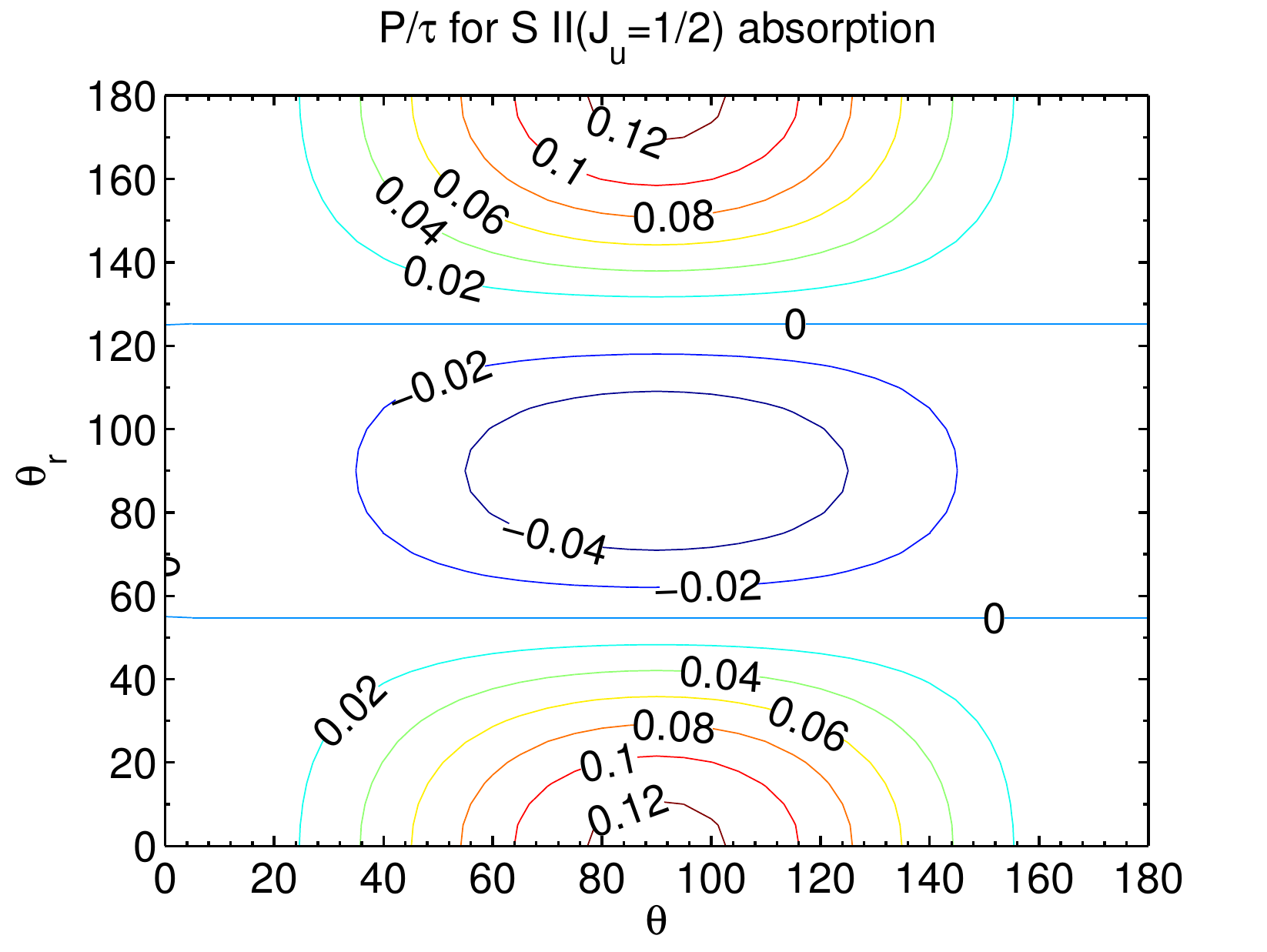}
\caption{{\it left}: Degree of polarization of S II absorption lines vs. $\theta_r$, the angle between magnetic field and direction of pumping source  $\theta$, the angle between magnetic field and line of sight.  {\it right}: The contour graphs of S II polarization. It is determined by the dipole component of density matrix $\sigma^2_0(\theta_r)$ and the direction of observation $\theta$ (Eq.\ref{absorb}). From \cite{YLfine}.}
\label{SII}
\end{figure}

Fig.\ref{SII} shows the dependence of polarization of S II absorption on $\theta_r, \,\theta$, which is representative for all polarization of absorption lines. From these plots, a few general features can be identified.  At $\theta=90^o$, the observed polarization reaches a maximum for the same $\theta_r$ and alignment, which is also expected from Eq.(\ref{absorb}). This shows that atoms are indeed realigned with respect to magnetic field so that the intensity difference is maximized parallel and perpendicular to the field (Fig.\ref{nzplane}{\it right}). At $\theta=0^\circ, 180^\circ$, the absorption polarization is zero according to Eq.(\ref{absorb}). Physically this is because the precession around the magnetic field makes no difference in the $x,y$ direction when the magnetic field is along the line of sight (Fig.\ref{nzplane}{\it right}). 

We consider a general case where the pumping source does not coincide with the object whose absorption is measured. 
If the radiation that we measure is also the radiation that aligns the atoms, the direction of pumping source coincides with line of sight, {\it i.e.}, $\theta=\theta_r$ (Fig.\ref{nzplane}{\it right}).

\subsection{Is there any circular polarization?}

Note that if the incident light is polarized in a different direction with alignment,
{\it circular polarization} can arise due to de-phasing though it is a 2nd order effect. Consider a background source with a nonzero Stokes parameter $U_0$ shining upon atoms aligned in $Q$ direction\footnote{To remind our readers, The Stokes parameters Q represents the linear polarization along ${\bf e}_1$ minus the linear polarization along ${\bf e}_2$; U refers to the polarization along $({\bf e_1+e_2})/\sqrt{2}$ minus the linear polarization along $({\bf -e_1+e_2})/\sqrt{2}$ (see Fig.\ref{nzplane}{\it right}).}. The polarization will be precessing around the direction of alignment and generate a $V$ component representing a circular polarization 
\be
\frac{V}{I\tau}\simeq \frac{\kappa_Q}{\eta_I}\frac{U_0}{I_0}=\frac{\psi_\nu}{\xi_\nu}\frac{\eta_Q}{\eta_I}\frac{U_0}{I_0}
\ee
where $\kappa$ is the dispersion coefficient, associated with the real part of the refractory index, whose imaginary part corresponds to the absorption coefficient $\eta$. $\psi$ is the dispersion profile and $\xi$ is the absorption profile.

The incident light can be polarized in the source, e.g. synchrotron emission from pulsars, or polarized through the 
propagation in the interstellar medium, e.g. as a result of selective extinction from the aligned dust grains (see
\cite{Lazarian07rev}). In the latter case, the polarization of the impinging light is usually low and the intensity of
circular polarization is expected to be low as well. This should not preclude the detection of the effect as the
instrumentation improves.

\section{IR/submilimetre transitions within ground state}

The alignment on the ground state affects not only the optical (or UV) transitions to the excited state, but also the magnetic dipole transitions within the ground state. Similar to HI, other species that has a structure within the ground state is also influenced by the optical pumping\footnote{To clarify, we do not distinguish between pumping by optical lines or UV lines, and name them simply "optical pumping".} through the Wouthuysen-Field effect (\cite{Wouthuysen:1952ij},\cite{Field58}). After absorption of a Ly$\alpha$, the hydrogen atom can relax to either of the two hyperfine levels of the ground state, which can induce a HI 21cm emission if the atom falls onto the ground triplet state (see also \cite{Furlanetto:2006hc} for a review). Recently, the oxygen pumping has been proposed as a probe for the intergalactic metals at the epoch of reionization \cite{Hernandez-Monteagudo:2007bs}. 

However, in all these studies, the pumping light is assumed to be isotropic. This is problematic, particularly for the metal lines whose optical depth is small. During the early epoch of reionization, for instance, the ionization sources are localized, which can introduce substantial anisotropy. The GSA introduced by the anisotropy of the radiation field can play an important role in many circumstances. The earlier oversimplified approach can lead to a substantial error to the predictions. The emissivity and absorption coefficients for the Stokes parameters are modified due to the alignment effect. The ratio of corresponding optical depth to that without alignment is
\bea
\tilde\tau&=&\frac{\tau}{\tau_0}=\frac{\tilde\eta-\tilde\eta_{s,i}\exp(-T_*/T_s)}{1-\exp(-T_*/T_s)},
\label{deltatau}
\eea
where $T_*$ is the equivalent temperature of the energy separation of the metastable and ground level, $T_s$ is the spin temperature.
\be
\tilde\eta_i=\eta_1/\eta_1^0=1+w_{0l}\sigma^2_0(J^0_l){\cal J}^2_0(i,\Omega),
\label{tdeta}
\ee 
\be
\tilde\eta_{s,i}=\tilde\epsilon_i =\epsilon_i/\epsilon^0_i=1+w_{l0}\sigma^2_0(J_l){\cal J}^2_0(i,\Omega) 
\label{tdetas}
\ee
are the ratios of absorption and stimulated emission coefficients with and without alignment, where ${\cal J}^2_0(i,\Omega)$ are the irreducible tensors for Stokes parameters \cite[see, e.g.][]{YLfine}. Since both the upper (metastable) level and the lower level are long lived and can be realigned by the weak magnetic field in diffuse medium, the polarization of {\em both emission and absorption} between them is polarized either parallel or perpendicular to the magnetic field like the case of all the absorptions from the ground state, and can be described by Eq.(\ref{absorb}). In the case of emission, the dipole component of the density matrix in  Eq.(\ref{absorb}) should be replaced by that of the metastable level. In fig.\ref{CI_OI}, we show an example of our calculation of the polarization of [CI] 610$\mu m$, which can be detected in places like PDRs.

We discussed pumping of hyperfine lines [H I] 21 cm and [N V] 70.7 mm in \cite{YLhyf} and fine line [O I] $63.2\mu m$ here. Certainly this effect widely exists in all atoms with some structure on ground state, e.g., Na I, K I, fine structure lines, [C I], [C II], [Si II], [N II], [N III], [O II], [O III],[S II], [S III], [S IV], [Fe II], etc (see Table~4.1 in \cite{Lequeux:2005tw}). The example lines we have calculated are listed in Table\ref{M1}. Many atomic radio lines are affected in the same way and they can be utilized to study the physical conditions, especially in the early universe: abundances, the extent of reionization through the anisotropy (or localization) of the optical pumping sources, and {\em magnetic fields}, etc. 

\begin{table*}
\begin{tabular}{ccccc}
\hline\hline
Lines&Lower level&Upper level&Wavelength ($\mu m$)&$P_{max}$\\
\hline
$[C I]$&$3P_0$&$3P_1$&$610$&20\%\\
$[O I]$&$3P_2$&$3P_1$&$63.2$&24\%\\
$[C II]$&$3P_{1/2}$&$3P_{3/2}$&$157.7$&2.7\%\\
$[Si II]$&$3P_{1/2}$&$3P_{3/2}$&$34.8$&4\%\\
$[S IV]$&$3P_{1/2}$&$3P_{3/2}$&$610$&10.5\%\\
\hline\hline
\end{tabular}
\caption{The polarization of forbidden lines.}
\label{M1}
\end{table*} 

\section{Atoms with hyperfine structure}

\begin{figure}
\includegraphics[%
  width=0.9\textwidth,
  height=0.28\textheight]{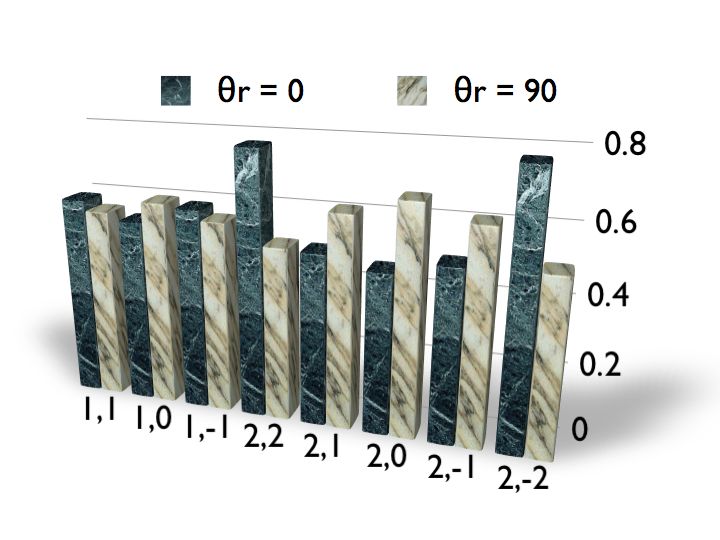}
  \caption{The occupation in ground sublevels for Na. It is modulated by the angle between magnetic field and the radiation field $\theta_r$. }
\label{Na}
\end{figure}

Although the energy of hyperfine interaction is negligible, the hyperfine interaction should be accounted for atoms with nuclear spins. This is because angular momentum instead of  energy is the determinative factor.  Hyperfine coupling increases the total angular momentum and the effects are two-sided. For species with fine structure ($J > 1/2$), the hyperfine interaction reduces the degree of alignment since in general the more complex the structure is, the less alignment is.  This is understandable. As polarized radiation is mostly from the sublevels with largest axial angular momentum, which constitutes less percentage in atoms with more sublevels. For species without fine structure ($J < 1/2$) like alkali, the hyperfine interaction enables alignment by inducing more sublevels\footnote{There are no energy splittings among them, the effect is only to provide more projections of angular momentum (see \cite{YLhyf}).} (see Fig.\ref{Na}).  

Note that the alignment mechanism of alkali is different from that illustrated in the carton (depopulation pumping) since the excitation rates from different sublevels on the ground state are equal. In other words, atoms from all the sublevels have equal probabilities to absorb photons. For the same reason, the absorption from alkali is not polarized\footnote{Only if hyperfine structure can be resolved, polarization can occur.}. The actual alignment is due to another mechanism, repopulation pumping. The alkali atoms are repopulated as a result of spontaneous decay from a polarized upper level (see \cite{Happer:1972ij}). Upper level becomes polarized because of differential absorption rates to the levels. 

\section{Polarization of resonance and fluorescence lines}

When the magnetic precession rate becomes less than the emission rate of the upper level, the direct effect of magnetic field on the upper level is negligible. The only influence of magnetic field is on the ground state through the alignment of atoms. This is the effect that was the focus of our studies in \cite{YLfine,YLhyf,YLHanle}. The atoms are aligned either parallel or perpendicular to the magnetic field as we discussed before. 

The differential occupation on the ground state (see Fig.\ref{Na}) can be transferred to the upper level of an atom by excitation. 
Emission from such a differentially populated state is polarized, the corresponding emission coefficients of the Stokes parameters are given in \cite{Landi-DeglInnocenti:1984kl, YLHanle} for fine structure transitions and \cite{YLhyf} for transitions involving hyperfine structures.
Indeed emission line can be polarized without GSA. This corresponds the textbook description of  polarization of scattered lines.  It can be shown that if the dipole and other higher order components of the ground state density matrix $\rho^2_{q'}$ are zeros, we recover the classical result in the optically thin case
\be
\epsilon_2=\epsilon_3=0,\, P =\frac{\epsilon_1}{\epsilon_0}= \frac{3E_1\sin^2\alpha}{4-E_1+3E_1\cos^2\alpha}
\label{classical_P}
\ee
where $\alpha$ is the scattering angle\footnote{Since there is no alignment on the ground state and we can choose the direction of radiation as the quantization axis, $\alpha=\theta$.}. The polarizability is actually given by 
\be
E_1=\frac{w^2_{J_uJ_l}r_{20}}{p_0}
\ee
for transitions in fine structure.
If we account for the alignment by radiation, but ignore magnetic field, {\em ie.}, $\theta_r=0$, then
\be
E_1=\frac{w^2_{J_uJ_l}\left[r_{20}+r_{22}\sigma^2_0(J_l)+\sqrt{2}p_2\sigma^2_0(J_l)\right]}{\sqrt{2}p_0+r_{02}\sigma^2_0}
\label{no_mag}
\ee

For optically thin case,  the linear polarization degree $p=\sqrt{Q^2+U^2}/I=\sqrt{\epsilon_2^2+\epsilon_1^2}/\epsilon_0$, the positional angle $\chi=\frac{1}{2}\tan^{-1}(U/Q)=\frac{1}{2}\tan^{-1}(\epsilon_2/\epsilon_1)$.

Since the weak magnetic field does not have direct influence on the upper level, there is no general simple geometrical correspondence between the polarization and magnetic field as in the case of absorption except the special case where the scattering angle is small (see \ref{B_effect} and \cite{Shangguan:2012} for details). In fact, from the discussions above (eq.\ref{classical_P}), we see that polarization is either parallel or perpendicular to the radiation field in the absence of GSA. The effect of GSA  is to introduce coherence on the upper level through the radiative excitation, which is then transferred to emission. To obtain the direction of magnetic field, one needs quantitative measurements of at least two lines. The lines for which we have calculated the polarizations are listed in Table \ref{emitable}.

\begin{table*}
\begin{tabular}{ccccc}
\hline\hline
Species&Lower state&Upper state&Wavelength(\AA)&$|P_{max}|$\\[0.5ex]
\hline
{S II}&$4S^o_{3/2}$&$4P_{3/2}$&1253.81&30.6\%\\
&&$4P_{5/2}$&1259.52&31.4\%\\
\hline
&$3P_{0}$&$3S^{o}$&1306&16\%\\
&$3P_{1}$&$3S^{o}$&1304&8.5\%\\
O I&$3P_{2}$&$3S^{o}$&1302&1.7\%\\
\cline{2-5} 
&$3P$&$3S^{o}$&5555,6046,7254&2.3\%\\
\cline{2-5}
&$3P_{0}$&$3D^o$&1028&4.29\%\\
&$3P_{1}$&$3D^o$&1027&7.7\%\\
&$3P_{2}$&$3D^o$&1025&10.6\%\\
\cline{2-5}
&$3P$&$3D^o$&5513,5958,7002&1.3\%\\
\hline
H I&
$1S_{1/2}$&
$2P_{3/2}$&
912-1216&26\%\\
\hline
{Na I}&
$1S_{1/2}$&
$2P_{3/2}$&
5891.6&20\%)\\
\hline
K I&
$1S_{1/2}$&
$2P_{3/2}$&
7667,4045.3&21\%\\
 \hline
{N V}&
  $1S_{1/2}$&
$2P_{3/2}$&
1238.8&22\%\\\\
 \hline
{P V}&
  $1S_{1/2}$&
$2P_{3/2}$&
1117.977&27\%\\
\hline
N I&
$4S^o_{3/2}$&
$4P_{1/2}$&
$1200$&5.5\%\\
\hline
Al II&$1S_0$&$1P^o_1$&8643\AA&20\%\\
\hline\hline
\end{tabular}
\caption{The maximum polarizations expected for a few example of emission lines and their corresponding transitions}
\label{emitable}
\end{table*}

\begin{figure}
\includegraphics[width=0.45\textwidth,
  height=0.28\textheight]{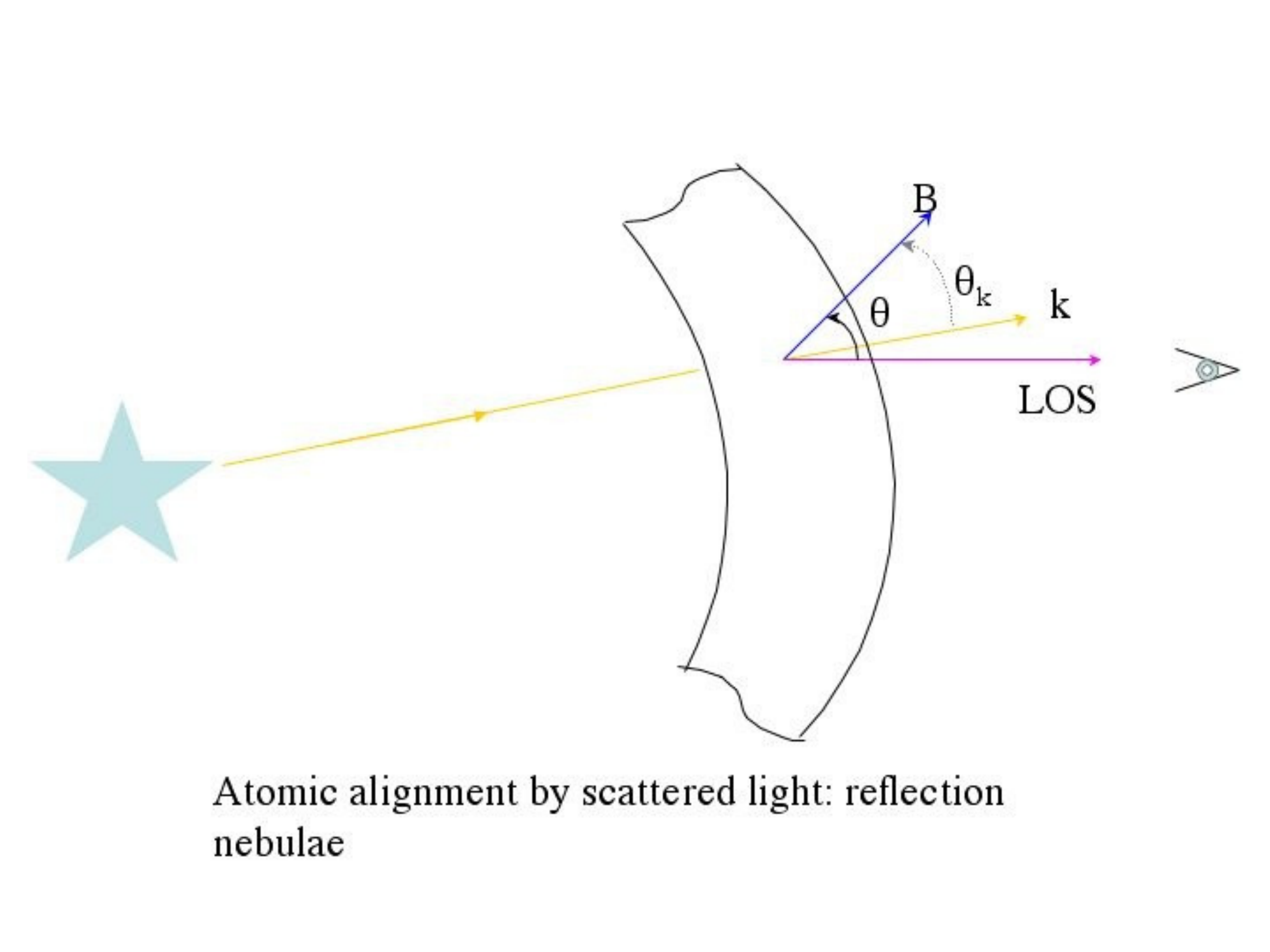}
\includegraphics[width=0.45\textwidth,
  height=0.28\textheight]{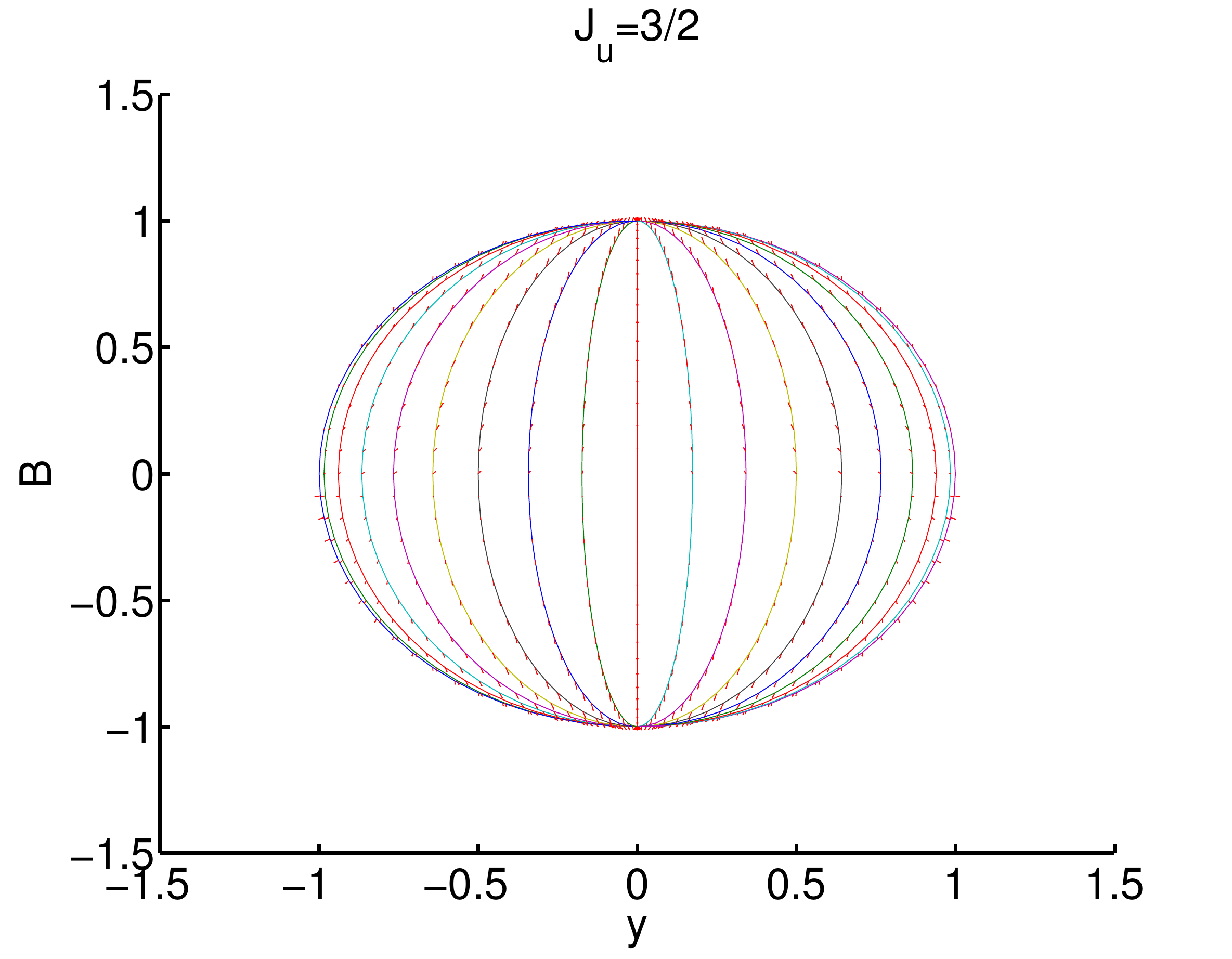}
  \caption{{\em left}: schematics of GSA by by circumstellar scattering; {\em right}: Polarization vectors of OI emission in a circumstellar region with alignment by uniform magnetic field. The inclination of magnetic field is 45 degree from the light of sight. The magnetic field is in the y direction in the plane of sky (from \cite{Yan:2009ys}).}
  \label{circumstellar}
\end{figure}

In the case that the direction of optical pumping is known, e.g., in planetary system and circumstellar regions, magnetic realignment can be identified if the polarization is neither perpendicular or parallel to the incident radiation (see eqs.\ref{classical_P},\ref{no_mag}). As an example, we show here polarization map from a spherical system with poloidal magnetic field \cite{Yan:2009ys}, eg., a circumstellar envelope, the polarization is supposed to be spherically symmetric without accounting for the effect of the magnetic field. With magnetic realignment though, the pattern of the polarization map is totally different, see fig.\ref{circumstellar}. {\em In practice, one can remove the uncertainty by measuring polarization from both alignable and non-alignable species, which does not trace the magnetic field. }

\section{Influence of Magnetic field on Polarization and line intensity} 
\label{B_effect}
\begin{figure}[!ht]
  \includegraphics[
  width=0.49\textwidth,
  height=0.3\textheight]{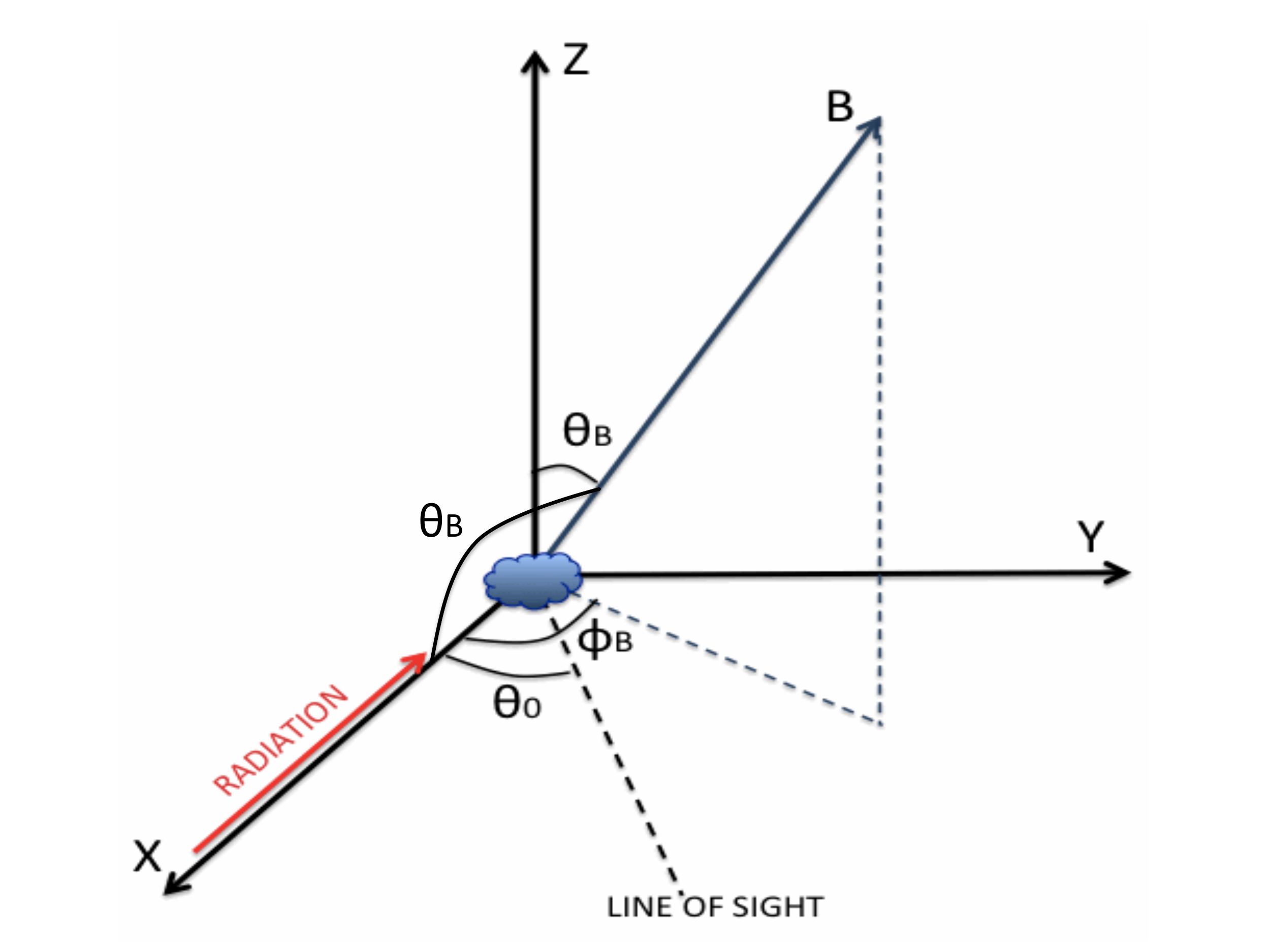}
   \includegraphics[
  width=0.49\textwidth,
  height=0.3\textheight]{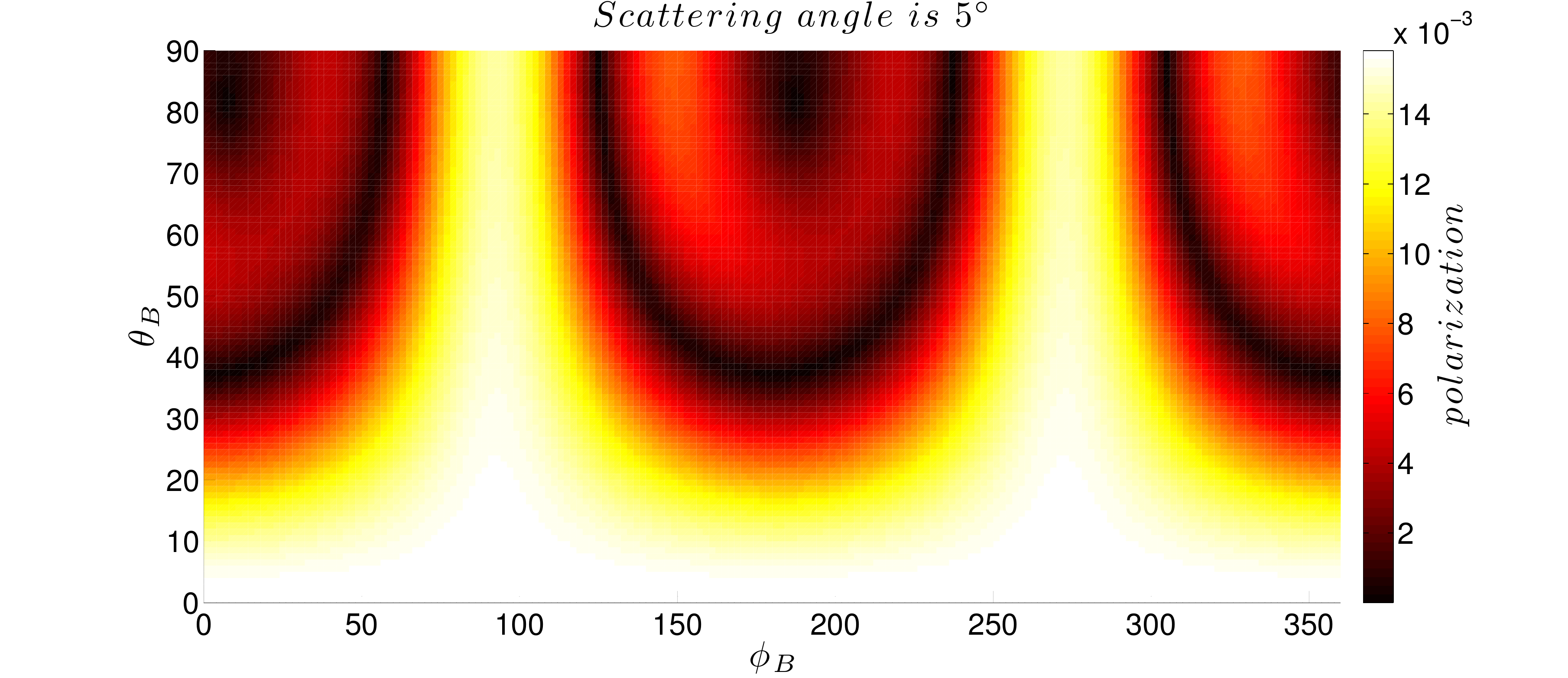}
  \caption{{\em Left}: the magnetic field ($\theta_B$, $\phi_B$) was measured in the reference frame where x-axis points to the radiation source, z-axis is perpendicular to the scattering plane. Scattering angle, $\theta_0$ is fixed; {\em Right}: The contour map of the polarization degree as a function of the direction of magnetic field. The scattering angle is fixed at $5^{\circ}$. ($\theta_{B}$, $\phi_{B}$) is the solid angle of the magnetic field s defined in Fig.\ref{rotate_dethe}. From \cite{Shangguan:2012}.}
  \label{rotate_dethe}
\end{figure}

The influence of magnetic field on the polarization of scattered lines can be illustrated by contour maps of position angle as a function of magnetic field direction for fixed scattering angle. The geometry of the maps is defined in Fig.\ref{rotate_dethe}{\em left}. 

\begin{figure}[!ht]
  \includegraphics[
  width=0.49\textwidth,
  height=0.3\textheight]{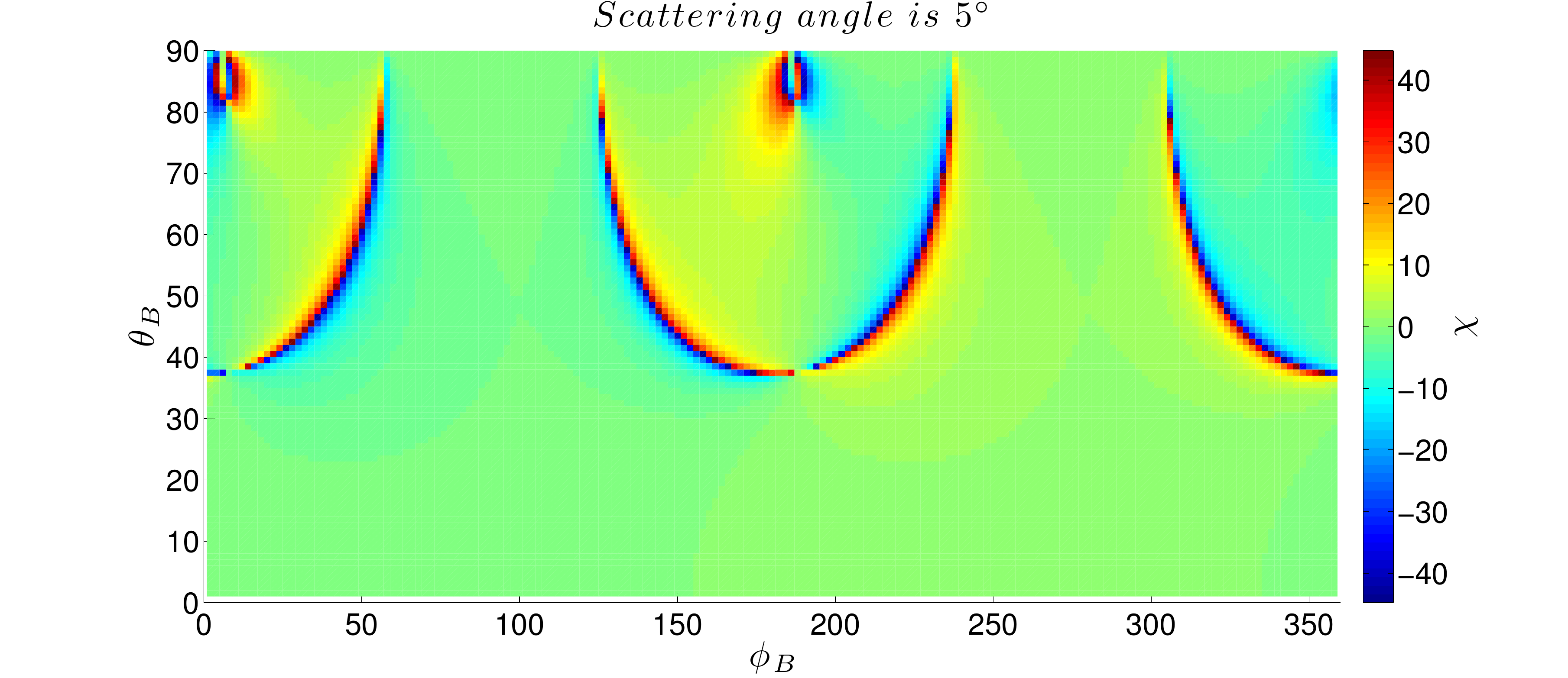}
  \includegraphics[
  width=0.49\textwidth,
  height=0.3\textheight]{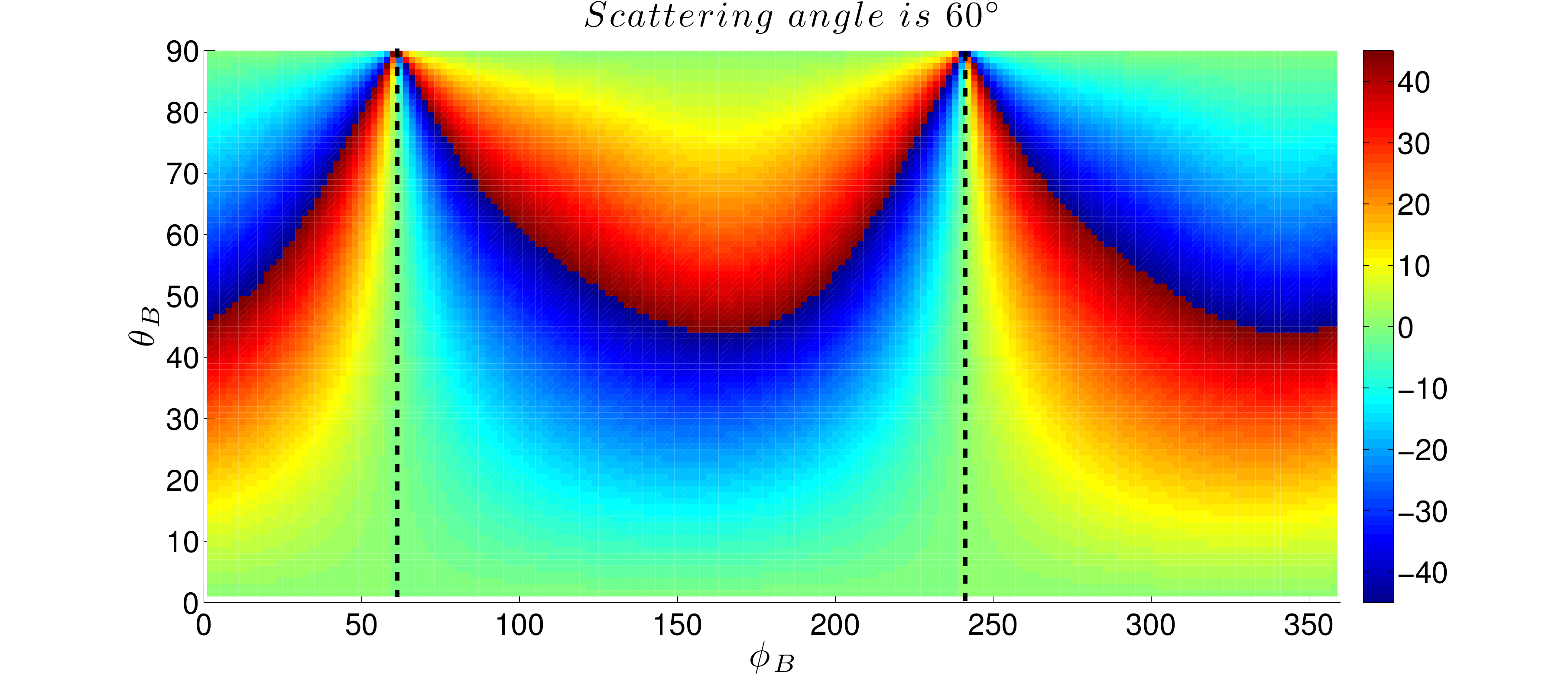}
  \caption{The contour map of position angle, $\chi$, as a function of the direction of magnetic field. The scattering angles are set to be $5^{\circ}$ ({\em left}) and $60^{\circ}$ ({\em right}). In the {\em right} panel, the dashed lines are two quasi-symmetry axes corresponding to $\phi_{B}=60^\circ$ and $240^\circ$, respectively. $\theta_{B}$ and $\phi_{B}$ are defined in Fig.\ref{rotate_dethe}. From \cite{Shangguan:2012}.}
  \label{chi_analysis_5}
\end{figure}

When scattering angle is relatively large ($\theta_0 \gtrsim 20^\circ$), quasi-symmetry is maintained and $\phi_B$ corresponding to the symmetric axes equals to the scattering angle (or to $\theta_0+180^\circ$) (see dash lines in Fig.\ref{chi_analysis_5}{\em right}). $\chi$ on the quasi-symmetric axes are zero. This means that when the projection of magnetic field on the scattering plane is parallel to the line of sight, the vector of polarization also lies in the same plane. In general, the position angle reduces with the decrease of the scattering angle. {\em When the scattering angle is small, the direction of polarization generally traces directly the magnetic field direction in the pictorial plane. (see Fig.\ref{chi_analysis_5}{\em left})} In addition, the polarizations are the same for $\theta_0$ and $(180^\circ-\theta_0)$  because of the symmetry. 

Magnetic field alters not only the direction of polarization of scattered lines, but also the degree of polarization (see Fig.\ref{rotate_dethe}{\em right}). Due to an average caused by magnetic mixing the magnetic field decreases the polarization degree at large scattering angle and increases the polarization at small scattering angle. In the classical description, there is some similarity to the Hanle process as discussed in \cite{Kastler1973, House1974}. The difference is that magnetic coherence is transferred to the upper state indirectly from the ground state through absorption \cite[see details in][]{Shangguan:2012}.

\section{Observational Perspective}
\subsection{Interstellar absorption and 3D magnetic field measurement}

Different from the case of emission, GSA is a unique mechanism to polarize absorption lines, which was first proposed by \cite{YLfine}. As illustrated above, 2D magnetic field in the plane of sky can be easily obtained by the direction of polarization. If we have quantitative measurement, we can get 3D magnetic field. 

The resonance absorption lines in the Table 2 appropriate for studying
magnetic fields in diffuse, low column density ($A_V \sim$ few tenths)
neutral clouds in the interstellar medium are NI, OI, SII, MnII, and
FeII. These are all in the ultraviolet.  

At higher column densities, the above lines become optically thick,
and lines of lower abundance, as well as excited states of the above
lines become available.  Significantly, some of these (TiII, FeI) are
in the visible.  An interesting region where the degenerate case
(pumping along the line of sight) should hold is the "Orion Veil"
\cite{Abel:2006tw}, a neutral cloud
with $A_V \sim 1.5$, and $N \sim 1000$, which is 1-2 parsec in the
foreground of the Orion Nebula. The Orion Veil should be pumped by the
Trapezium.  This region is of particular interest for magnetic field
studies, since it is one of the only places where maps exist of the HI
21 cm Zeeman Effect.  This gives the sign and magnitude of the
magnetic field along the line of sight.  The magnetic realignment
diagnostic, on the other hand, gives the orientation of the magnetic
field lines in 3D, which is exactly complimentary information:
combination of the two yields a complete 3D magnetic field map.

\begin{figure}
\includegraphics[%
  width=0.32\textwidth,
  height=0.2\textheight]{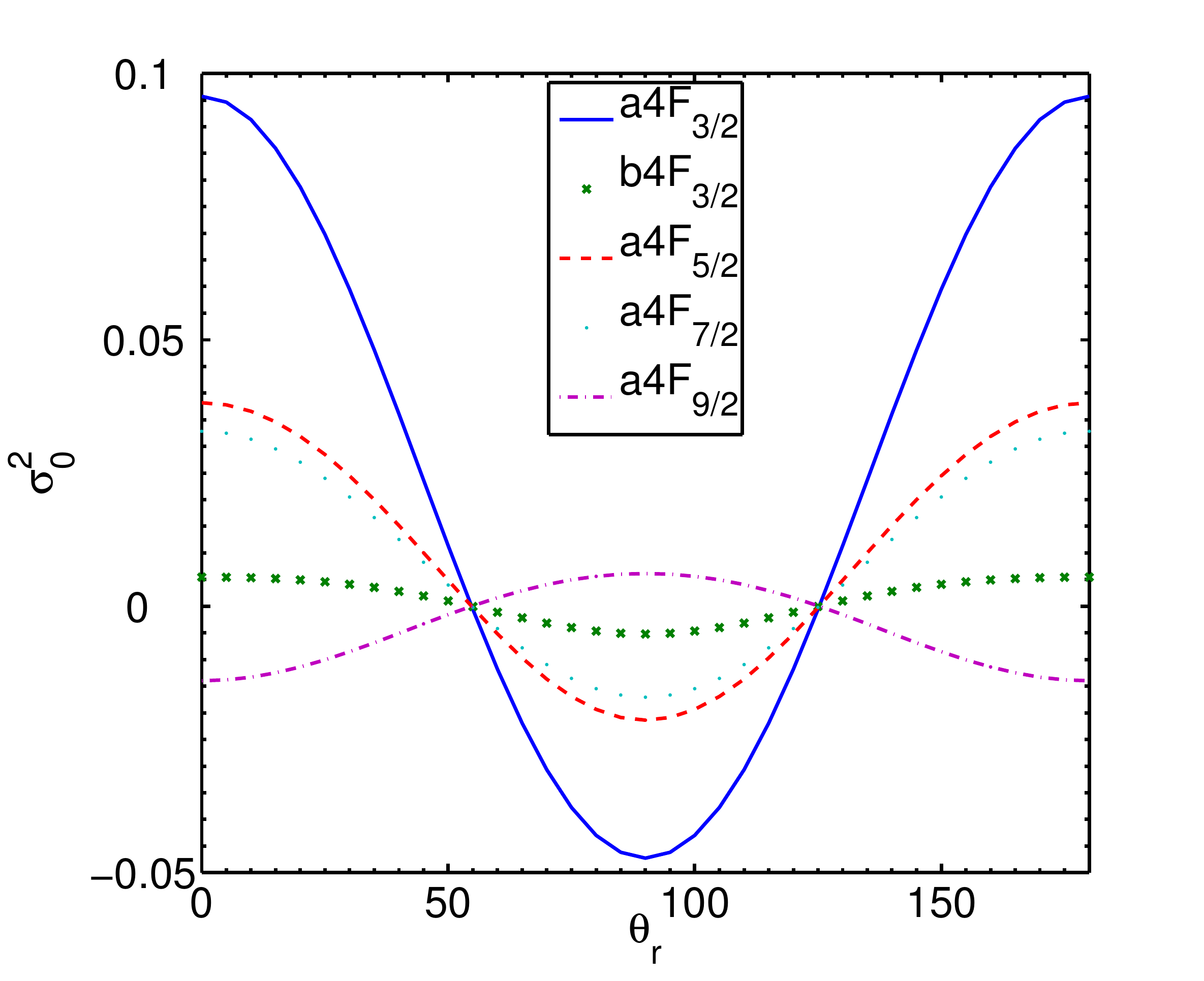}
  \includegraphics[%
  width=0.32\textwidth,
  height=0.2\textheight]{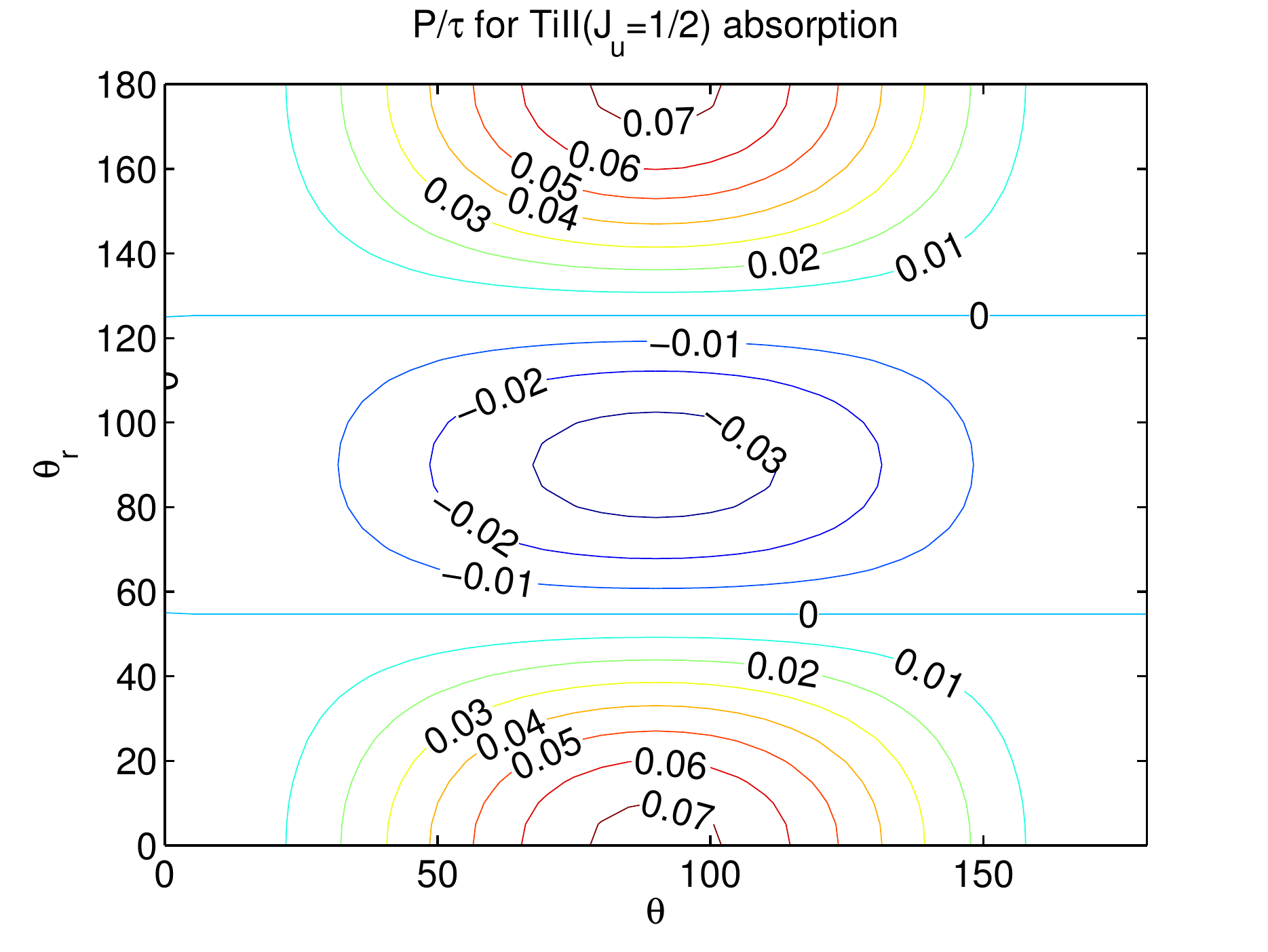}
\includegraphics[%
  width=0.32\textwidth,
  height=0.2\textheight]{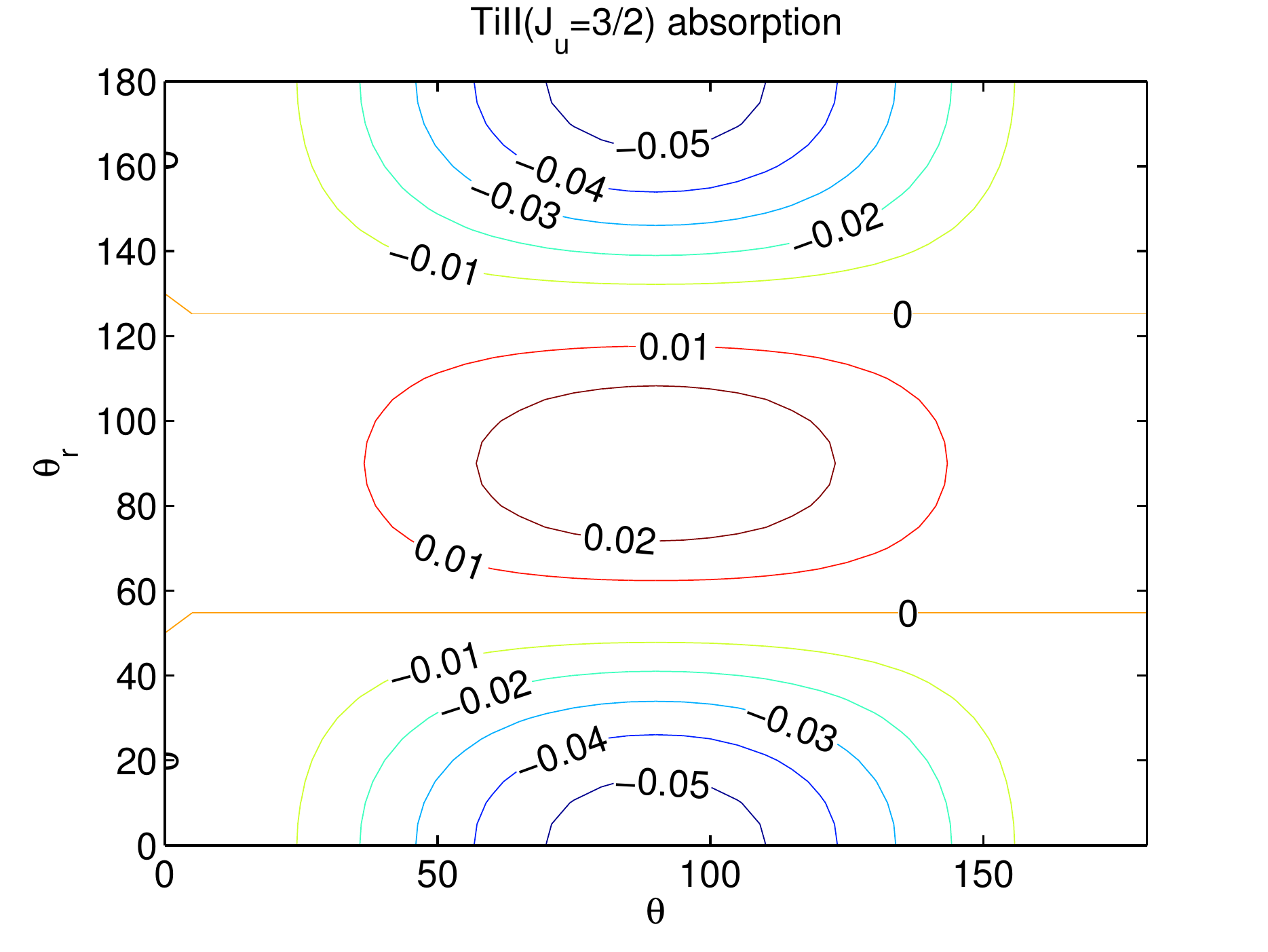}
  \caption{{\em left}: the alignment of the ground state $a4F$ and metastable level $b4F_{3/2}$ of Ti II; {\em middle} and {\em right}: the contour of equal degree of polarization of Ti II absorption lines ($J_l=1/2 \rightarrow J_u=1/2,3/2$). $\theta_r$ and $\theta$ are the angles of incident radiation and line of sight from the magnetic field (see Fig.\ref{regimes}{\em right}). In the case of pumping source coincident with the background source, we have the degeneracy and polarization will be determined by one parameter $\theta_r=\theta$. From \cite{YLHanle}}
\label{TiII}
\end{figure}

The TiII lines provide a fairly accessible test of the magnetic
alignment diagnostic, although there are not enough strongly polarized lines in the visible
to be useful by themselves: observation of an effect at TiII with
moderate resolution would motivate a serious study of the pumping of
the much more numerous FeI lines, and the construction of more
advanced instrumentation.  Table 3 shows the five strongest
TiII lines, all from the J = 3/2 ground state of the ion (see also Fig.\ref{TiII}).  The
polarization depends only on the J of the upper state, regardless of
the other quantum numbers.  We have assumed "weak pumping" for this
calculation, so that the rate of decay from the excited states of the
ground term exceeds the pumping rate.  The most strongly polarized
line, 3074, is unfortunately difficult from the ground, and the
strongest line, 3384, is only weakly polarized.  The 3243 line is the
most favorable. It is estimated that for a line with $\tau \sim 1$ and a
width of $20$ km/sec, this effect would be detectable at 10$\sigma$
for all stars brighter than V=8 with the spectropolarimeter at SALT
with resolution R = 6000 (Nordsieck, private communications).
  
As a first step, with low resolution measurement, 2D magnetic field 
in the pictorial plane can be easily obtained from the direction of polarization with a 90 degree degeneracy, 
similar to the case of  grain alignment and Goldreich-Kylafis effect. Different from the case of emission, 
any polarization, if detected, in absorption lines, would be an exclusive indicator of alignment, and it traces magnetic field 
since no other mechanisms can induce polarization in absorption lines. The polarization in H$\alpha$ absorption that \cite{Kuhn:2007vn}
 reported in fact was due to the GSA as predicted in \cite{YLfine} for general absorptions. 

If we have two measurable,  we can solve Eq.\ref{absorb} and obtain $\theta_r,\, \theta$. With $\theta_r$ known, the 90 degree degeneracy can be removed since we can decide whether the polarization is parallel or perpendicular to the magnetic field in the plane of sky.  Combined with $\theta$, the angle between magnetic field and line of sight, we get 3D direction of magnetic field.       

\subsection{Circumstellar Absorption}

The Be star $\zeta$ Tau illustrates one application of the diagnostic in circumstellar matter.
A number of absorption lines from SiII metastable level $2P_{3/2}$ are seen, along with the OI
and SII lines already discussed.  This provides for a large number of
different species.  In this case the absorption is likely being formed
in a disk atmosphere, absorbing light from the disk and pumped by
light from the star. The net polarization signal is integrated across
the disk, and depends just on the magnetic field geometry and the
inclination of the disk.  In
this case, the inclination is known from continuum polarization
studies to be $79^o$ \cite{Carciofi:2005bh}. The FUSP sounding rocket should be
sensitive to these effects: $\zeta$ Tau is one of the brightest UV
sources in the sky.

A second interesting circumstellar matter case is for planetary disks
around pre-main sequence stars.  In this case, pumping conditions are
similar to those for comets in the Solar System: pumping rates on the
order of $0.1 - 1$ Hz, and realignment for fields greater than $10 -
100$ mGauss.  Conditions here are apparently conducive to substantial
populations in CNO metastable levels above the ground term: Roberge,
et al (2002) find strong absorption in the FUV lines (1000 - 1500 Ang)
of OI (1D) and NI and SII (2D), apparently due to dissociation of
common ice molecules in these disks (these are also common in comet
comae).  Since these all have $J_l >1$, they should be pumped, and
realigned.  This presents the exciting possibility of detecting the
magnetic geometry in planetary disks and monitoring them with time.
Since these are substantially fainter sources, this will require a
satellite facility.
                            
\subsection{Fluorescence from reflection nebulae}

The magnetic realignment diagnostic can also be used in fluorescent scattering lines.  This is because the
alignment of the ground state is partially transferred to the upper
state in the absorption process.  There are a number of fluorescent
lines in emission nebulae that are potential candidates (see \cite{Nordsieck:2008kx}). Although such
lines have been seen in the visible in HII regions (\cite{Grandi:1975cr,Grandi:1975nx})
and planetary nebulae \cite{Sharpee:2004oq}, we suggest that
reflection nebulae would be a better place to test the diagnostic,
since the lack of ionizing flux limits the number of levels being
pumped, and especially since common fluorescent ions like NI and OI
would not be ionized, eliminating confusing recombination radiation. Realignment should make itself evident in a line polarization whose
position angle is not perpendicular to the direction to the central
star.  This deviation depends on the magnetic geometry and the
scattering angle.  The degree of polarization also depends on these
two things. It will be necessary to compare the polarization of
several species with different dependence on these factors to separate
the effects.  This situation is an excellent motivation for a pilot
observation project.

\subsection{Magnetic field in PDR regions}

Most fine structure FIR lines arise from photon dominated (PDR) region, a transition region between fully ionized and molecular clouds illuminated by a stellar source of UV radiation. The line ratio of the brightest ones, e.g., [C II] $158\mu m$, [O I] $63,~145\mu m$, are used to infer physical parameters, including density and UV intensity based on the assumption that they are collisionally populated. Recent observations of UV absorption by Sterling et al. \cite{Sterling:2005fv}, however, find that the population ratio of $3P_{1,0}$, the originating levels of [O I] $63,~145\mu m$ is about twice the LTE value in the planetary nebula (PN) SwSt 1, and fluorescence excitation by stellar continuum is concluded to be the dominant excitation mechanism. In this case, the alignment is bound to happen on the two excited levels $3P_{1,0}$ because of the anisotropy of the pumping radiation field, resulting polarizations in the [O I] $63,~145\mu m$ lines.

The high spatial resolution of SOFIA, for instance, is advantageous in zooming into PDRs and resolving the lines. Moreover, in highly turbulent environment, we expect that the magnetic field is entangled. The higher resolution means less averaging in the signals from atoms aligned with the magnetic field. The high sensitivity of the upgraded HAWC++ also provides us a possibility of doing precise quantitative measurement of the spectral polarizations, which can resolve the ambiguity of the 90 degree degeneracy and enables a 3D topology of magnetic field, which cannot be obtained from any other present magnetic diagnostics.   

\section{Observational testings}
 
 Here we demonstrate a synthetic observation of polarization of the sodium D lines  can reveal the magnetic fields of Jupiter and the solar wind that interacts with the comet via magnetic alignment. The advantage of direct studies of magnetic perturbations by spacecrafts has been explored through many important missions.  However, the spacecraft measurement are rather expensive. Are there any other cost-effective ways to study magnetic turbulence in interplanetary medium?

Comets are known to have sodium tails and sodium is an atom that can be aligned by radiation and realigned by solar wind magnetic fields. This opens an opportunity of studying magnetic fields in the solar wind from the ground, by tracing the polarization of the Sodium line.
We simulated the
the degree of polarization and its variations arising from resonant scattering from Na line in a magnetized comet wake \cite{YLhyf, Yan09}.  The study has been extended for comet Halley and Jupiter's Io based on the magnetic field data from Vega1\&2 during the encounter with comet Halley in 1986 and the Galileo's mission to Jupiter, respectively \cite[see details in][]{Shangguan:2012}. 

Though the abundance of sodium in comets is very low, its high efficiency in
 scattering sunlight makes it a good tracer \cite{Thomas:1992kh}. As discussed in \cite{YLhyf}, the gaseous sodium atoms in the comet's tail acquire angular momentum from the solar radiation, i.e. they are aligned. Resonant scattering from these aligned atoms is polarized.  Collision rates in the coma of comets are very low, less than once/day beyond a few 1000 km from the nucleus, so that fluorescent alignment should be rapidly established in the outflowing sodium from the comet \cite{Harris:1997}.
 
\begin{figure*}
\includegraphics[width=.44\textwidth]{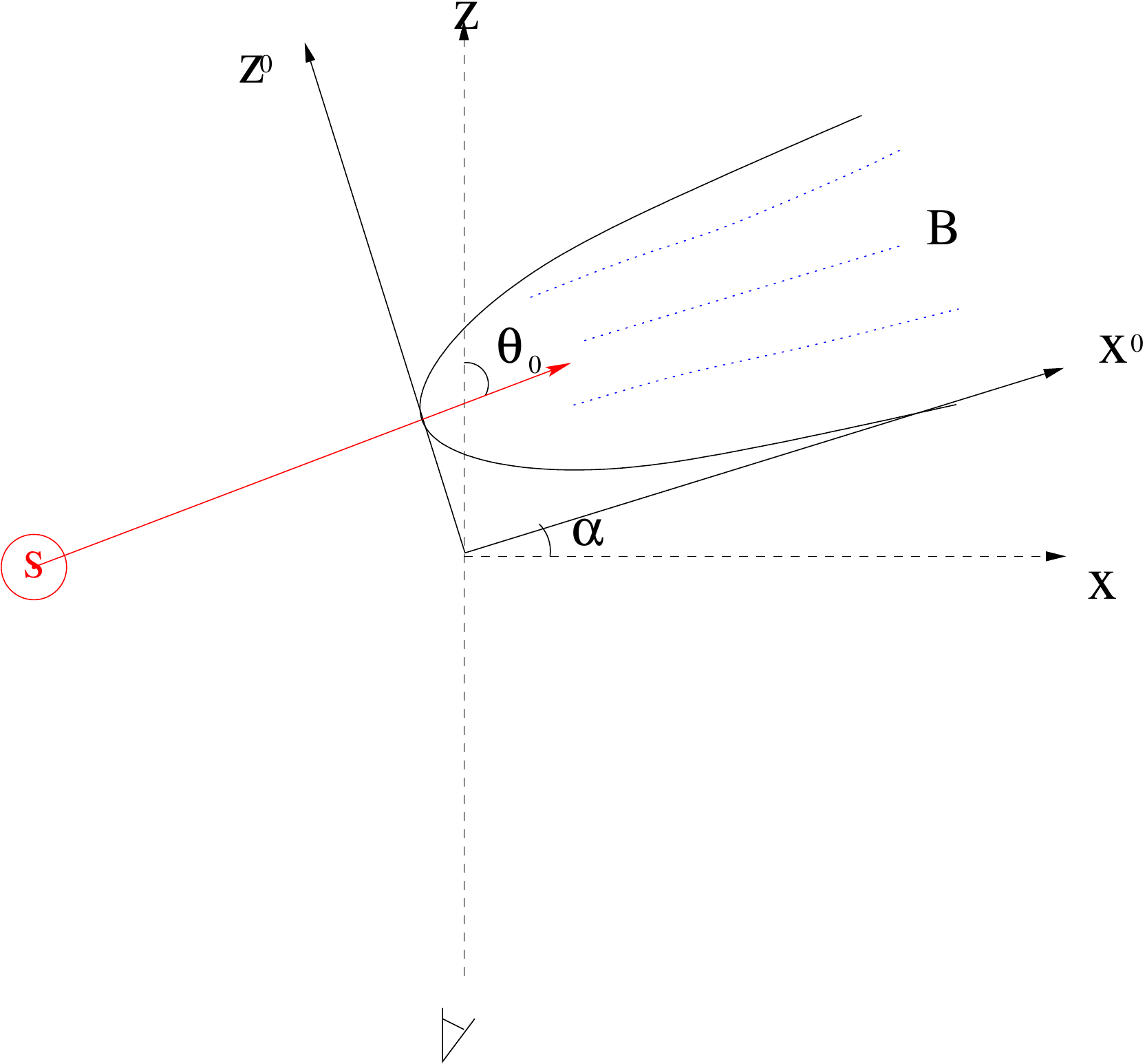}
\includegraphics[width=.54\textwidth]{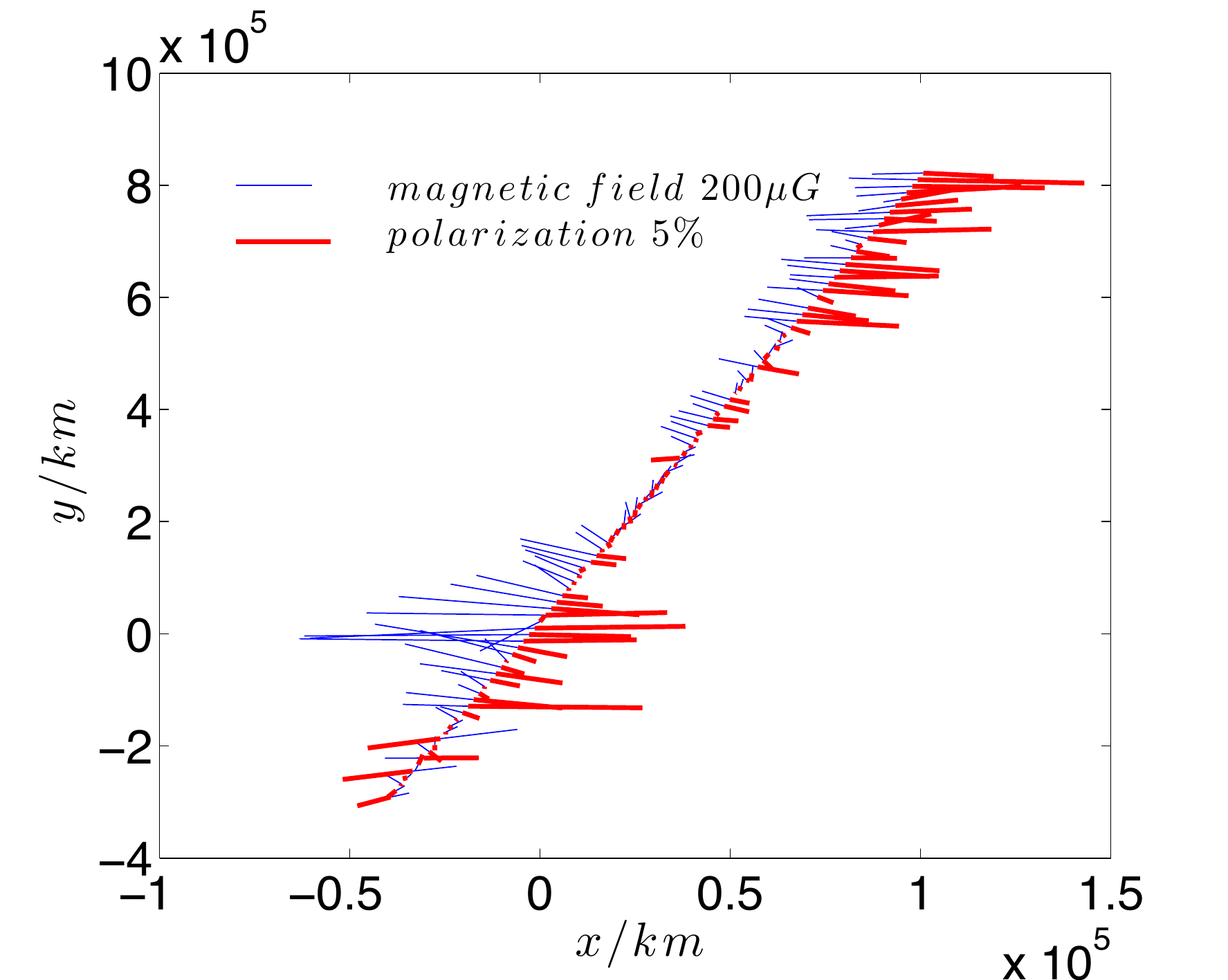}
\caption{{\it Left}: Schematics of the resonance scattering of sunlight by the sodium in comet wake. The sodium tail points in the direction opposite to the Sun. The observer on the Earth sees the stream at the angle $\theta_0$. Magnetic field
realigns atoms via fast Larmor precession. Thus the polarization traces the interplanetary magnetic fields. {\it Right}:  the magnetic field and polarization along the trajectory of Vega1 encountering the comet Halley (from \cite{Shangguan:2012}).}
\label{cometmag}
\end{figure*}

 As shown in Fig. \ref{cometmag}, the geometry of the 
scattering is well defined, i.e., the scattering angle $\theta_0$ is known. The alignment is modulated by the local magnetic field. The polarization of the sodium emission thus provides exclusive 
information on the magnetic field in the interplanetary medium. Depending on its 
direction, the embedded magnetic field alters the degree of 
alignment and therefore the polarization of the light scattered by the aligned atoms. Fig.\ref{cometmag} illustrates the trajectory of a comet along which the magnetic field varies and the polarization of Sodium D2 emission changes accordingly. By comparing observations with them, we can cross-check our model and determine the structure of magnetic field in the heliosphere with similar observations. One can investigate not only spatial, but also {\em temporal} variations
of magnetic fields. Since alignment happens at a time scale $\tau_R$, magnetic field variations on this time scale will be reflected. This can allow for a cost-effective way of studying
interplanetary magnetic turbulence at different scales.

GSA provides us a unique opportunity to detect the magnetic field that is beyond the reach of space probes. The polarization direction traces directly the orientation of magnetic field, especially, when the scattering angle is small. We can acquire the entire accurate information of the magnetic field with two or more lines from alignable species. Since alignment happens on a very short timescale (inverse of photo-excitation rate), instantaneous tomography of the magnetic field can be realized through GSA.

\section{GSA and Cosmological studies}

\subsection{Magnetic field in the epoch of reionization?}

The issue of magnetic field at the epoch of reionization is a subject of controversies. The fact that the levels of O I ground state can be aligned through anisotropic pumping suggest us a possibility of using GSA to diagnose whether magnetic field exists at that early epoch. 

The degree of polarization in the optically thin case can be obtained in a similar way as above by replacing $\tilde\eta_0, \tilde\epsilon_0$ by $\tilde\eta_1, \tilde\epsilon_1$. In the alignment regime, the Stokes parameters, U=0, and therefore,
\bea
P=\frac{Q_\nu}{B_\nu(T_{CMB})}&=&\tau\left\{\frac{\tilde\epsilon_1[\exp(T_*/T_{CMB})-1]}{\tilde\eta_1 \exp(T_*/T_s)-\tilde\epsilon_1}-1\right\}\nonumber\\
&\simeq&\tau_0[\tilde\epsilon_1(1+y_{iso}/\tau_0)-\frac{\tilde\eta_1-\tilde\epsilon_1}{1-\exp(-T_*/T_s)}]\nonumber\\
&=&\tilde\epsilon_1 y_{iso}-\tau_0\frac{1.5\sin^2\theta [w_{l0}\sigma^2_0(J_l)-w_{0l}\sigma^2_0(J^0_l)]/\sqrt{2} }{1-\exp(-T_*/T_s)}
\label{Povertau}
\eea

\begin{figure} 
\includegraphics[width=0.32\textwidth,
  height=0.2\textheight]{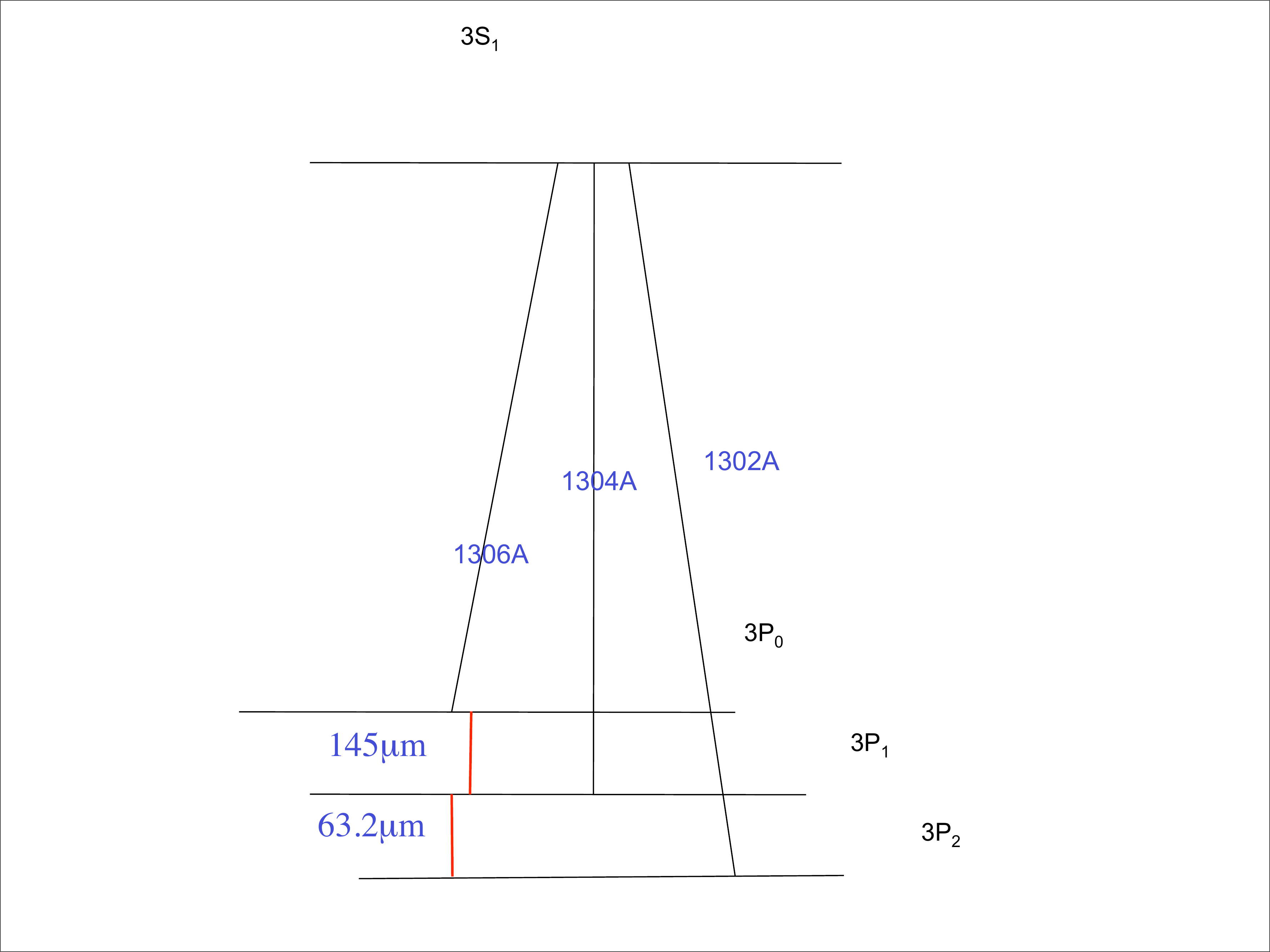}
\includegraphics[width=0.32\textwidth,
  height=0.2\textheight]{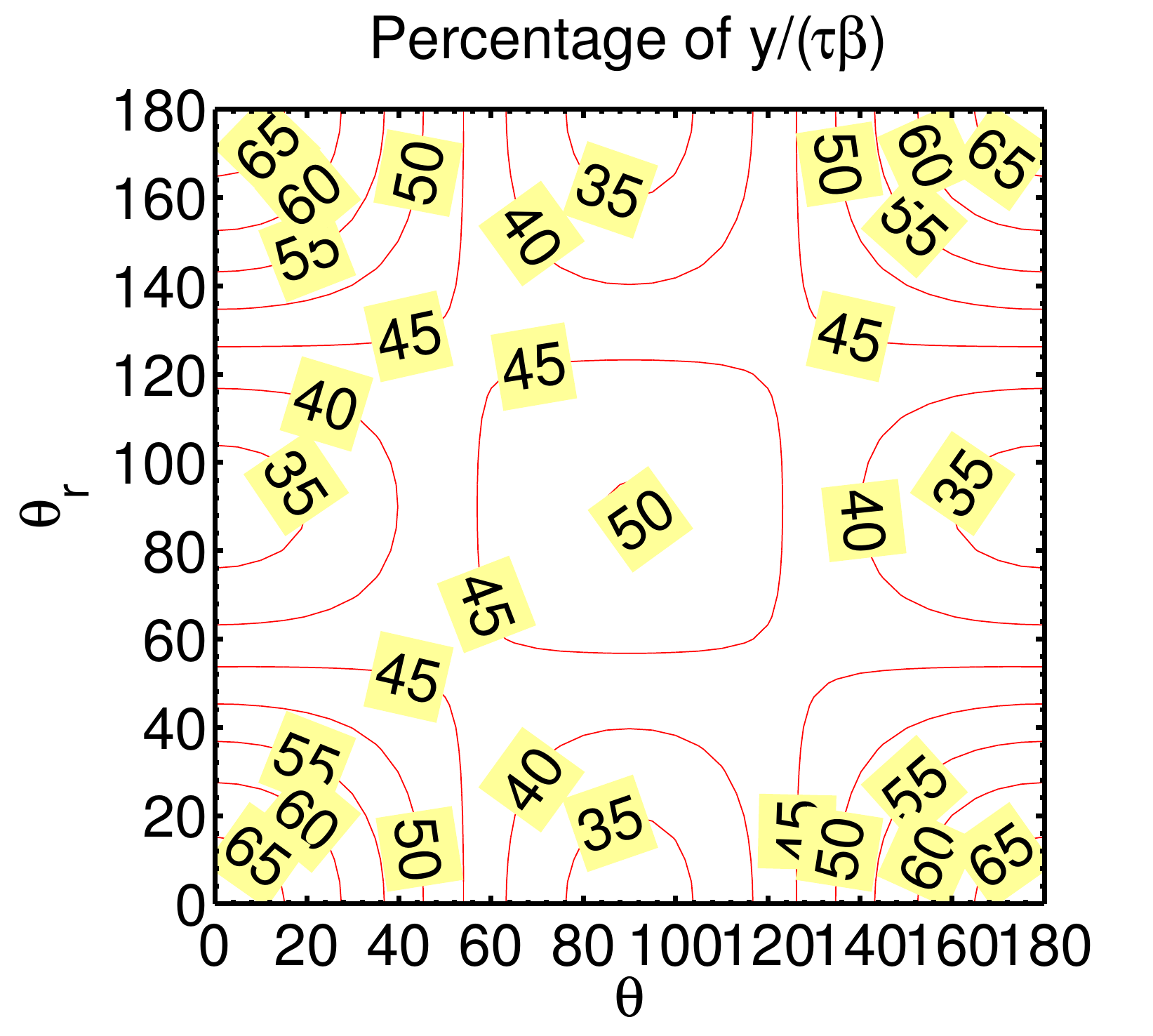}
  \includegraphics[width=0.32\textwidth,
  height=0.2\textheight]{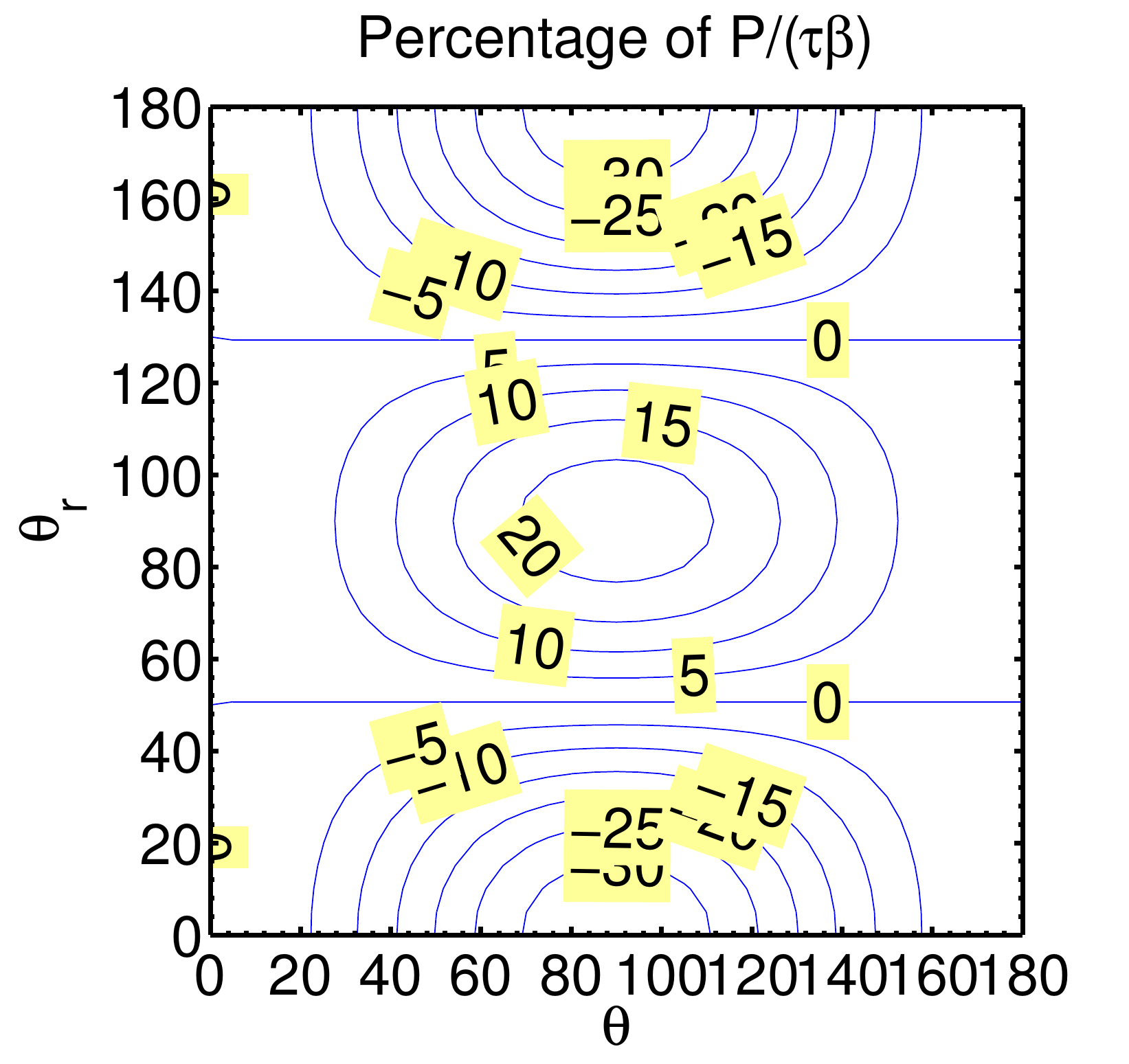}
\caption{ {\em Left}: the the schematics of fluorescence pumping of [O I] line; {\em middle \& right}: $y/(\tau\beta)$, $P/(\tau\beta)$ of [OI] line in early universe. $\theta_r,\, \theta$ are respectively the angles of the incident radiation and l.o.s. from the magnetic field. From YL08.}
\label{CI_OI}
\end{figure}

In the case of nonzero magnetic field, the density matrices are determined by $\theta_r$, the angle between magnetic field as well as the parameter $\beta$. Similar to $y$, the degree of polarization is also proportional to $\beta$. In Fig.\ref{CI_OI}, we show the dependence of the ratios $y/(\tau\beta)$, $P/(\tau\beta)$ on $\theta_r$ and $\theta$. Since U=0, the line is polarized either parallel ($P>0$) or perpendicular ($P<0$) to the magnetic field. The switch between the two cases happen at $\theta_r=\theta_V=54.7^o, 180-54.7^o$, which is a common feature of polarization from aligned level (see \cite{YLfine,YLHanle} for detailed discussions). 

\subsection{Influence on abundance studies}

\begin{figure}[!ht]  
  \includegraphics[
  width=0.45\textwidth,
  height=0.3\textheight]{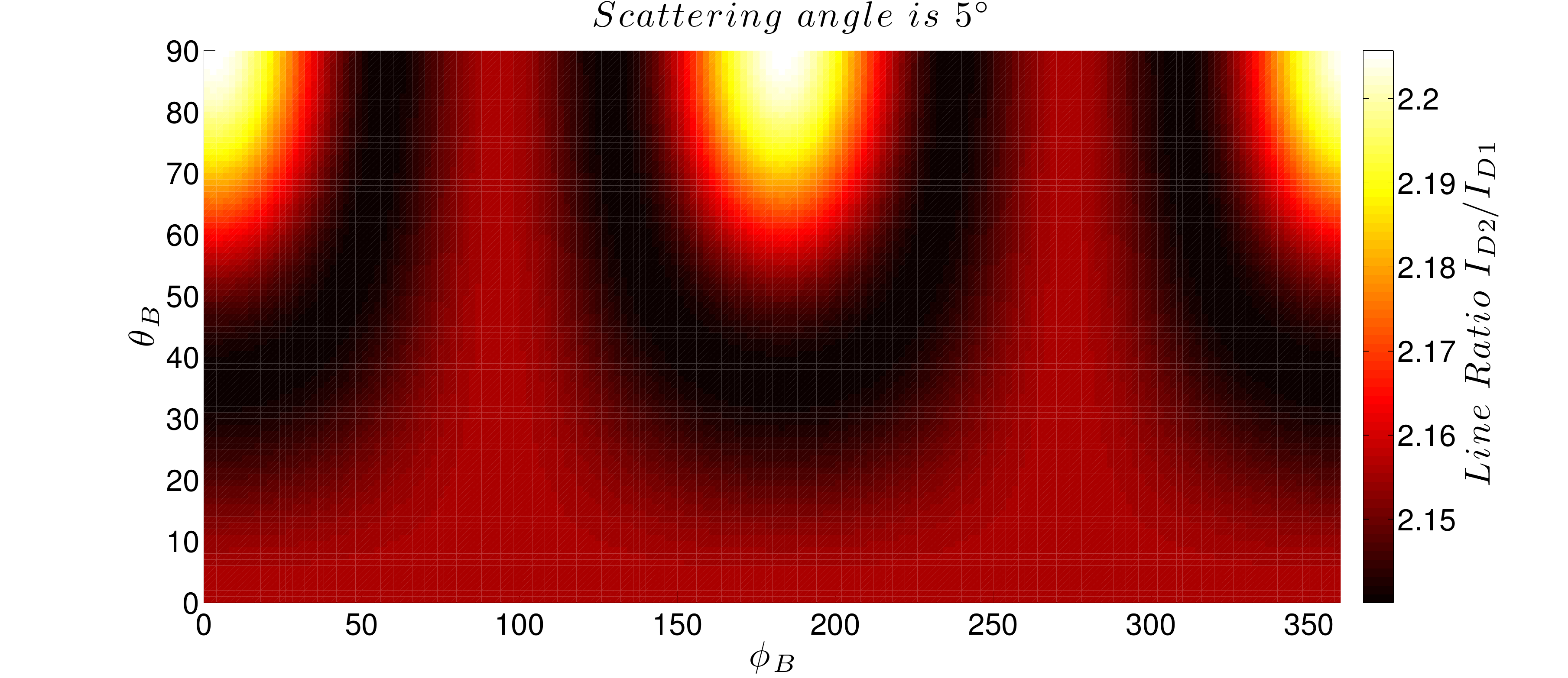}  
\caption{The contour map of line ratio, $I_{D2}/I_{D1}$ as a function of the direction of magnetic field. The scattering angle is also set to be $5^{\circ}$. From \cite{Shangguan:2012}.}
\label{lineratio}
\end{figure}

The GSA not only induces/influences the polarization of spectral lines, but also modulate the line intensity. This can cause substantial error in the estimates of chemical abundances if this effect is not included. This has been shown in last section on the pumping of [OI] line in early universe. The same also happens with the permitted absorption and emission lines (see Fig.\ref{lineratio}).

\subsection{Metal detection in early universe}  
For instance, the distortion of CMB due to the optical pumping calculated accounting for the anisotropy of the optical/UV radiation field for the optically thin case differs from the result without anisotropy included \cite{Yan:2009ys}:

\bea
y=\frac{\Delta I_\nu}{B_\nu(T_{CMB})}&=&\tau\left\{\frac{\tilde\epsilon_1[\exp(T_*/T_{CMB})-1]}{\tilde\eta \exp(T_*/T_s)-\tilde\eta_s}-1\right\}\nonumber\\
&=&\tau_0\left[\frac{\tilde\epsilon_1[\exp(T_*/T_{CMB})-1]}{\exp(T_*/T_s)-1}-\tilde\tau
\label{deltaI}\right]\nonumber\\
&\simeq&\tau_0[\tilde\epsilon_1(1+y_{iso}/\tau_0)-\tilde\tau]\nonumber\\
&=&\tilde\epsilon_1 y_{iso}+\tau_0\frac{[w_{l0}\sigma^2_0(J_l)-w_{0l}\sigma^2_0(J^0_l)](1-1.5\sin^2\theta)/\sqrt{2}}{1-\exp(-T_*/T_s)}
\label{yovertau}
\eea
where $y_{iso}$ is the distortion neglecting the anisotropy of the radiation field and GSA (see \cite{Hernandez-Monteagudo:2007bs}). Indeed if alignment is not accounted, then 
\be
y=y_{iso}=\tau_0\frac{T_*}{T_s}\frac{\Delta T}{T_{cmb}}=\tau_0\frac{T_{cmb}}{T_s}\exp\left(\frac{T_*}{T_{cmb}}\right)\left[1-\frac{A(J^0_l)}{\sum_{J_l} A(J_l)}\right]\frac{[J^0_l]\beta B_mI_m}{[J_l](A_m+B^s_mI_m)},
\ee
where $A_m, B_m, B^s_m$ are the Einstein coefficients for the magnetic dipole transitions within the ground state, and $I_m$ is the corresponding line intensity, $\beta\equiv BI_\nu/B_mI_m$. Both $\tilde\eta_s$ and $\tilde\tau$ depend on the line of sight and the GSA (Eqs.\ref{deltatau}, \ref{tdetas}), the resulting distortion in radiation is thus determined by the angle $\theta$ as well as the UV intensity of the OI line $I_\nu$ (or $\beta$, see Fig.\ref{CI_OI}). Since both of the two terms on the right hand side of Eq.\ref{yovertau} are proportional to $\beta$, the resulting distortion $y$ is also proportional to $\beta$. 

In some sense, this study is made for the case of weak pumping regimes discussed in \cite{YLfine}. But we take into account in addition the absorption and stimulated emission within the ground state.

\section{Feasibility of GSA studies}

The GSA is a subtle effect that is described by a complex quantum electrodynamics formalism. Therefore it is important to stress that this does not translates into the effect being difficult to measure. Indeed, as we mentioned earlier, the effect was studied successfully in the laboratory many years ago. The predicted polarization measures, as we shown in the text, can be very substantial. Moreover, different species show different degrees of alignment (including zero alignment) and this allows to separate the GSA-induced polarization from the polarization of instrumental origin. 

We have also discussed that the effects related to the GSA, but in a different regime of magnetic saturation have been successfully studied in the solar atmosphere.This vividly supports our claim that the GSA effect is not only measurable in laboratory, but also in not pure astrophysical conditions\footnote{incidentally,
these studies induced a local revolution in understanding of the solar spectra. We expect even deeper impact of the GSA studies. Indeed, the domain of the applicability of the GSA is really extensive and the consequences of the magnetic field and abundance studies are extremely important.}. In fact, the polarization of Kuhn et al. (2007, \cite{Kuhn:2007vn}) circumstellar envelope already indicate the effect of the GSA polarization of absorption lines predicted in Yan \& Lazarian (2006, \cite{YLfine}). 

In terms of observational studies, one should remember that it usually takes some time for the technique to get accepted. One may recall a long history of attempts of measuring of the Goldreich-Kylafis effect. However, now the technique is routinely used. We feel that a similar process takes place with the practical usage of the GSA effect. We hope that this review can accelerate the observational work towards practical use of the GSA.

\section{Summary}

GSA is an important effect the potential of which for magnetic field studies has not been yet
tapped by the astrophysical community. The alignment itself is an effect studied well in the laboratory;
the effects arise due to the ability of atoms/ions with fine and hyperfine structure to
get aligned in the ground/metastable states. Due to the long life of the atoms in such states the Larmor
precession in the external magnetic field imprints the direction of the field into the polarization of emitting
and absorbing species. This provides a unique tool for studies of magnetic fields using polarimetry of UV, optical and radio
lines. The range of objects for studies is extremely wide and includes magnetic fields in the early universe,
in the interplanetary medium, in the interstellar medium, in circumstellar regions. Apart from this, the consequences
of alignment should be taken into account for correct determining the abundances of alignable species.

As astrophysical magnetic fields cover a large range of scales, it is important to have techniques to study magnetic
fields at different scales. In this respect atomic realignment fits a unique niche as it reveals small scale structure of magnetic field. For instance, we have discussed the possibility of studying magnetic fields in interplanetary medium.
This can be done without conventional expensive probes by studying polarization of spectral lines. In some cases
spreading of small amounts of sodium or other alignable species can produce detailed magnetic field maps of
a particular regions of interplanetary space, e.g. the Earth magnetosphere.

{\it Acknowledgments}. HY acknowledges the support from 985 grant from Peking University and the ``Beyond the Horizons'' grant from Templeton foundation as well as the visiting professorship at the International Institute of Physics 
(Brazil). AL's research is supported by the NSF AST 1109295 and the NSF Center for Magnetic 
Self-Organization (CMSO). He also acknowledges the Humboldt Award and related productive stay at the
Universities of Bochum and Cologne. 

{\small
\bibliography{yan}
\bibliographystyle{apj}
}
\end{document}